\documentclass[iop]{emulateapj}

\usepackage{xcolor}
\usepackage[normalem]{ulem}

\shorttitle{On the formation of a stable penumbra in a region of flux emergence}
\shortauthors{Murabito et al.}

\begin{document}


\title{On the formation of a stable penumbra in a region of flux emergence in the Sun}


\author{M. Murabito\altaffilmark{1}, P. Romano\altaffilmark{2}, S.L. Guglielmino\altaffilmark{1},  F. Zuccarello\altaffilmark{1}}
\email{mmurabito@oact.inaf.it}




\altaffiltext{1}{Dipartimento di Fisica e Astronomia - Sezione Astrofisica, Universit\`{a} degli Studi di Catania,
			 Via S. Sofia 78, 95123 Catania, Italy}
\altaffiltext{2}{INAF - Osservatorio Astrofisico di Catania,
              Via S. Sofia 78, 95123 Catania, Italy.}		 			 


\begin{abstract}

We studied the formation of the first penumbral sector around a pore in the following polarity of the Active Region (AR) NOAA 11490. We used a high spatial, spectral, and temporal resolution data set acquired by the Interferometric BIdimensional Spectrometer operating at the NSO/Dunn Solar Telescope as well as data taken by the Helioseismic and Magnetic Imager onboard the Solar Dynamics Observatory satellite. 
On the side towards the leading polarity, elongated granules in the photosphere and an arch filament system (AFS) in the chromosphere are present, while the magnetic field shows a sea-serpent configuration, indicating a region of magnetic flux emergence. 
We found that the formation of a stable penumbra in the following polarity of the AR begins in the area facing the opposite polarity located below the AFS in the flux emergence region, differently from what found by Schlichenmaier and colleaguestbf. Moreover, during the formation of the first penumbral sector, the area characterized by magnetic flux density larger than 900 G and the area of the umbra increase.      

\end{abstract}




\section{Introduction}

Magnetic flux emergence and its successive interaction with the magnetic canopy represent a crucial point in the first stage of the formation of the sunspot penumbrae. 


\citet{Sch10b} found that the penumbra forms in sectors and generally on the side opposite to that between the two polarities of an active region (AR). Also, they found that individual filaments form on the side towards the opposite polarity, but these are not stable and disappear on granular lifetimes. In this area, elongated granules are present. They continuously appear and disappear, indicating a region of flux emergence as observed in the numerical simulations of \citet{Cheung08} and \citet{Tort09}. \citet{Sch10b} concluded that a stable penumbra cannot form in the region of flux emergence, because the strong dynamics introduced by ongoing flux emergence prevent the settlement of the penumbral field.
 
Elongated granules in the photosphere are developed by emerging horizontal fields, being aligned with the horizontal component of the emerged field \citep{Strous99, Cheung07, Cheung08}.
Recently, \citet{Lim13} found that penumbral filaments were formed in a region on the side toward the opposite polarity of the AR. In this region a series of elongated granules was detected in association with an emerging flux region. However, they detected other cases where the flux emergence did not form the penumbra, even if accompanied by a series of elongated granules. They explained this behavior with the presence or absence of strong overlying canopy fields.

In this context, we recall that an important signature of magnetic flux emergence in the chromosphere is a bundle of dark arches crossing the polarity inversion line, called arch filament system \citep[AFS; e.g.,][]{Bruz80,Spad04,Zucc05}. Such a structure is usually observed in the hours following the appearance of H${\alpha}$ brightenings.

Here, we present one of the rare observations of the formation of the penumbra, starting from the side towards the opposite polarity, below an AFS, in the AR NOAA 11490. 
In the next Section we describe the whole data set and its analysis. In Section 3 we present the results. Finally, in Section 4 we summarize our conclusions.

\section{Observations and analysis}

We study the following spot of the AR NOAA 11490 using high temporal, spatial, and spectral resolution data acquired by the Interferometric BIdimensional Spectrometer \citep[IBIS;][]{Cav06} operating at the NSO/Dunn Solar Telescope (DST). The observations were carried out on 2012 May 28 from 13:39 UT to 14:38 UT.

The whole data set acquired on 2012 May 28, whose relevant characteristics were already described in detail in \citet{Rom13}, consists of 45 scans through the \ion{Fe}{1} 617.30 nm, \ion{Fe}{1} 630.25 nm, and \ion{Ca}{2} 854.2 nm lines, with 67 s cadence. Among these, 15 scans are relevant to the following part of the AR and were acquired from 14:19 UT to 14:38 UT, as shown in the Figure 1 of \citet{Rom13}. The observed fields of view (FOVs) are shown in Figure 1a. The lines were sampled with a spectral profile having a full with half maximum of 2 pm and an average wavelength step of 2 pm. 
The \ion{Fe}{1} 617.3 nm and 630.25 nm lines were sampled in spectro-polarimetric mode with 30 and 24 spectral points, respectively. The signal-to-noise ratio of these measurements is about $3 \times 10^{-3}$  in units of continuum intensity per Stokes parameter. The \ion{Ca}{2} 854.2 nm line was sampled in spectroscopic mode with 25 spectral points. The FOV of each camera was $500\times 1000$ pixels, with a pixel scale of 0\farcs09. 

\begin{figure*}[htbp]
	\centering
	\includegraphics[scale=0.35,clip, trim=80  70 40 200]{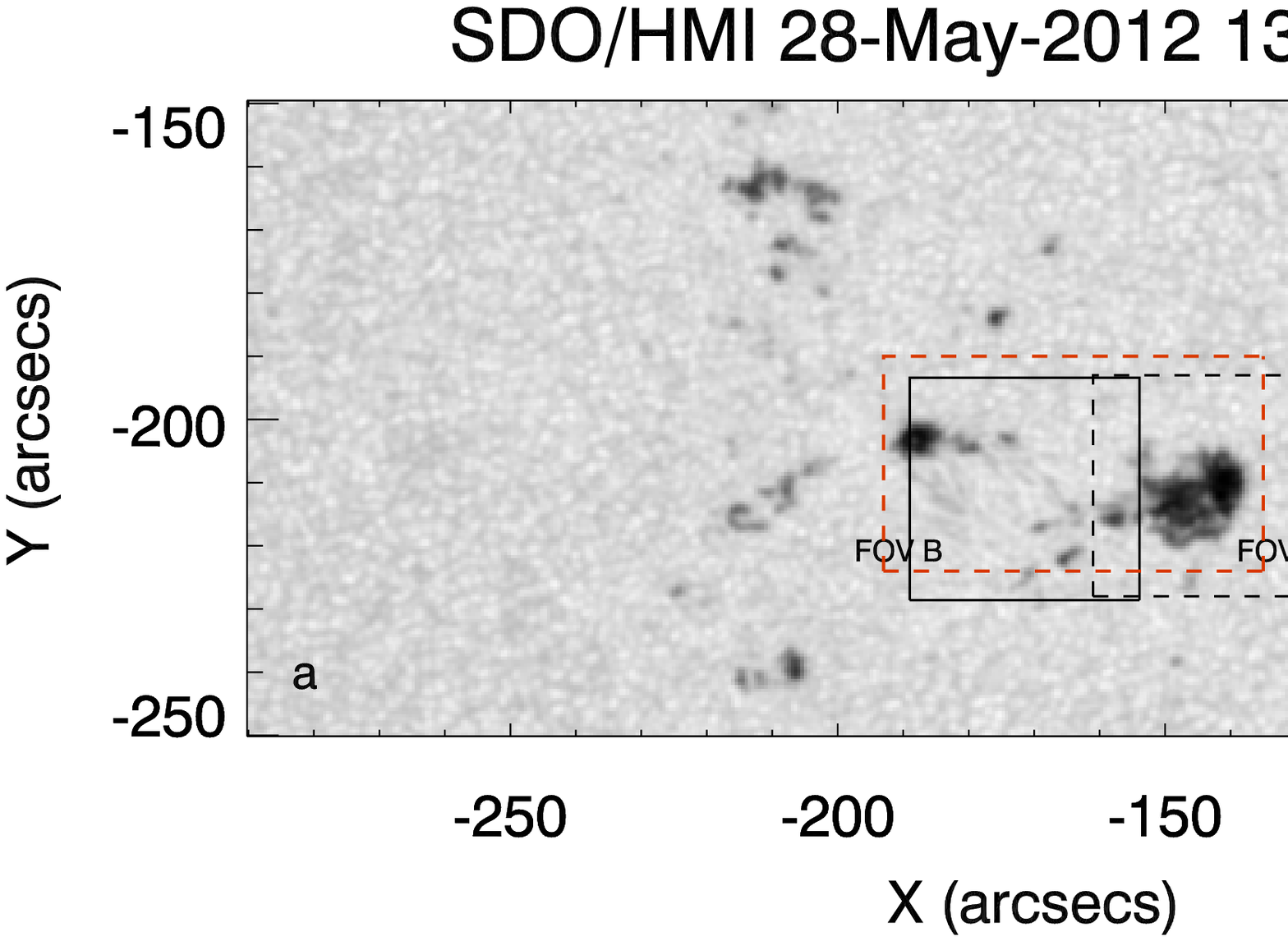}%
	\includegraphics[scale=0.35,clip, trim=80  70 40 200]{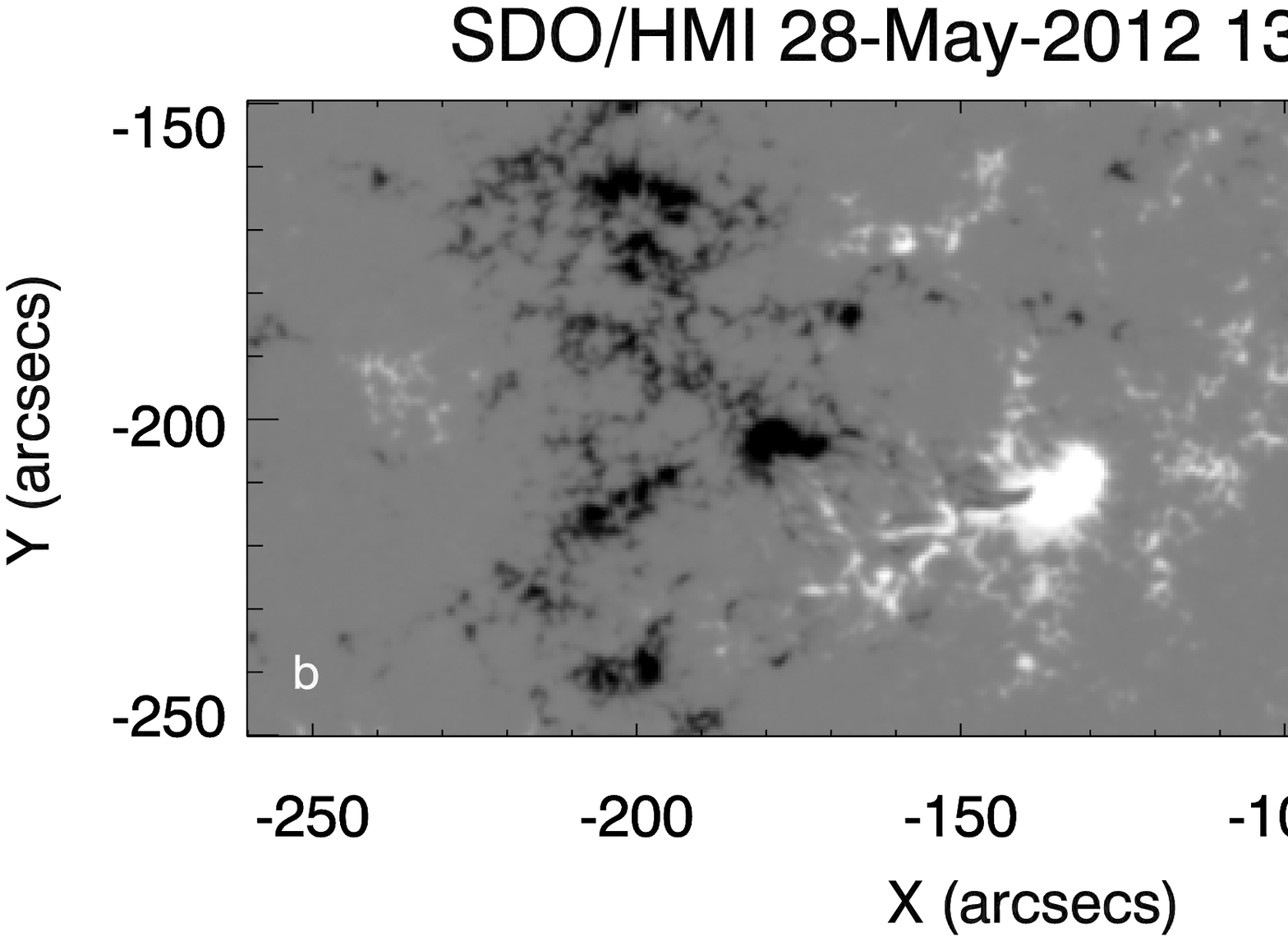}
	\includegraphics[scale=0.35,clip, trim=80  40 40 200]{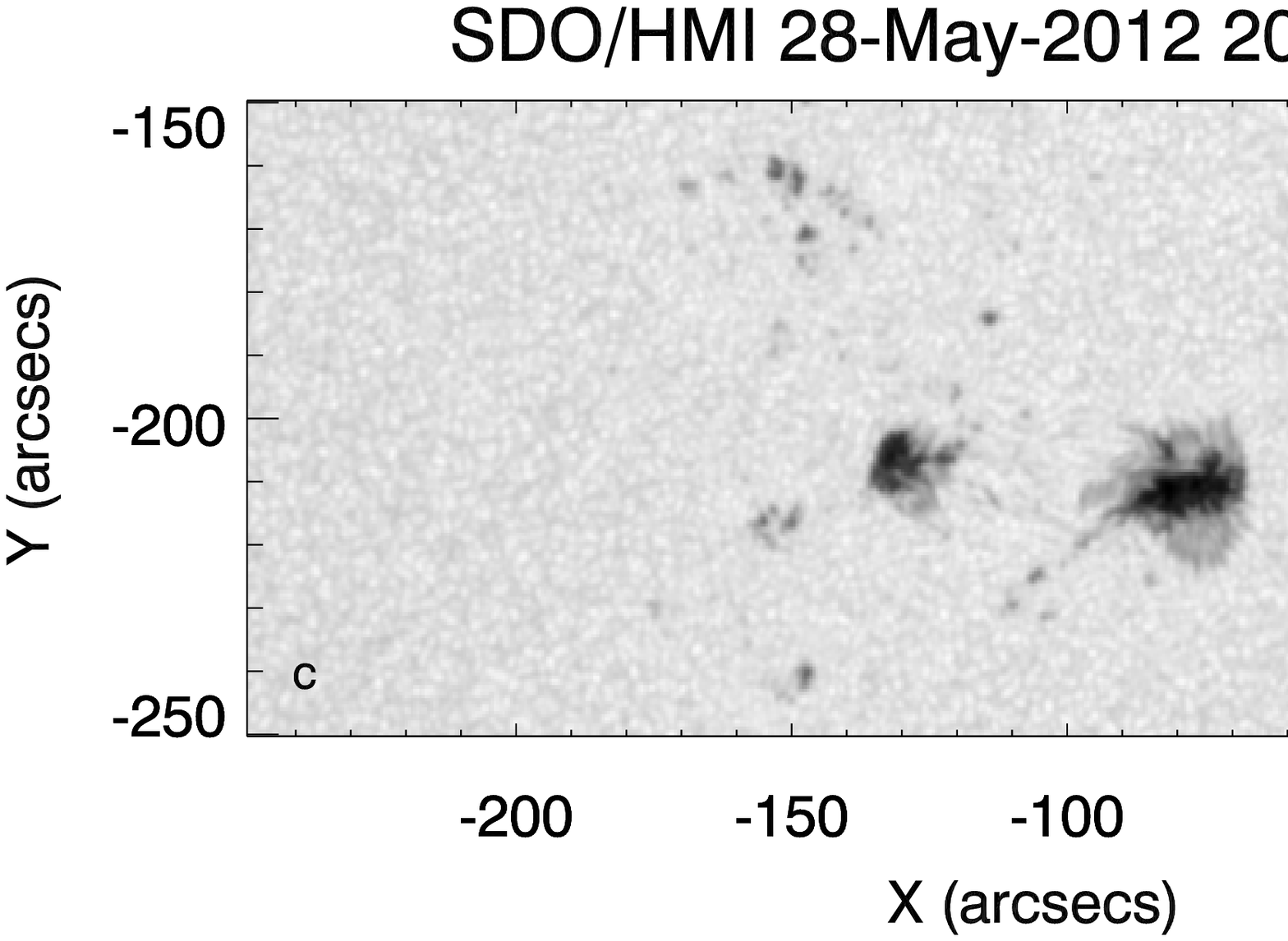}%
	\includegraphics[scale=0.35,clip, trim=80  40 40 200]{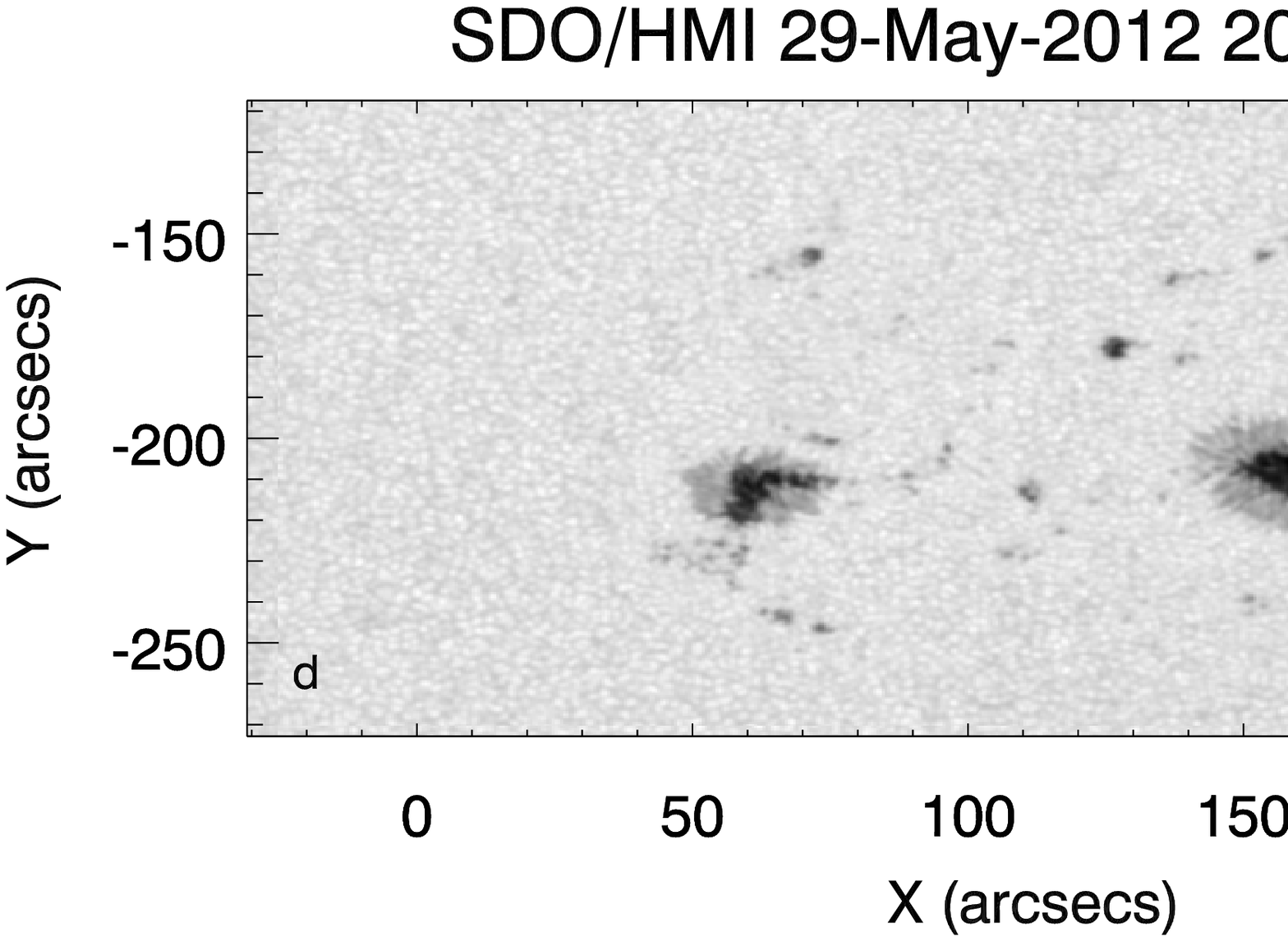}
	\caption{Continuum filtergrams (a-c-d) and LOS magnetogram (b) taken by SDO/HMI showing maps of AR NOAA 11490 at different times. The black dashed and continuum boxes in (a) indicate the IBIS FOVs containing the leading and the following spots, respectively, while the red box indicates the FOV shown in Figure 2. Here and in the following figures, North is at the top, West is to the right. The axes give the distance from solar disc center. The arrow points to the disc center.}
	\label{Figure 1}
\end{figure*}

For each spectral frame, a simultaneous broad-band (at $633.32 \pm 5$ nm) frame was acquired, imaging the same FOV observed in spectro-polarimetric mode and with the same exposure time. To reduce the seeing degradation, the images were restored using the Multi-Frame Blind Deconvolution \citep[MFBD;][]{Lof02} technique (see details in \citealp{Rom13}).

To determine the strength and the inclination angle of the magnetic field we performed a single-component inversion of the Stokes profiles for all the available scans of the \ion{Fe}{1} 617.3 nm line using the SIR code \citep{SIR}. The procedure was described in detail in a previous paper \citep{Mur16}.  

To study the evolution in the forming penumbra, we also analyzed both Space-weather HMI Active Region Patches \citep[SHARPs,][]{Bobra14} continuum filtergrams and magnetograms taken by the Helioseismic and Magnetic Imager (HMI) on the Solar Dynamics Observatory \citep[SDO,][]{Sch12} satellite. These data are taken in the \ion{Fe}{1} 617.3 nm line with a resolution of 1\arcsec. They cover seven hours of observations, starting from 2012 May 28 at 13:58:25 UT until 20:58:25 UT, with a cadence of 12 minutes.

IBIS and SDO/HMI observations were co-aligned using as a reference the first spectral image in the continuum of the \ion{Fe}{1} 617.3 nm line taken by IBIS at 14:20 UT on 2012 May 28 and the continuum filtergram closest in time taken by SDO/HMI. We used the IDL \textit{SolarSoft} mapping routines to take into account the different pixel sizes.

Maps of the mean circular polarization averaged over the line, $V_{s}$, and of the mean linear polarization signal, $L_{s}$, were calculated using 
\begin{eqnarray}
V_{s}=\frac{1}{12\left \langle I_{c} \right \rangle}\sum_{i=1}^{12} \epsilon_{i} V_{i} \\
L_{s}=\frac{1}{12\left \langle I_{c} \right \rangle}\sum_{i=1}^{12} \sqrt{Q^{2}_{i}+U^{2}_{i}}
\end{eqnarray} 
where $\left \langle I_{c} \right \rangle$ is the continuum intensity averaged over the IBIS FOV, \textit{i} runs over the 12 central wavelength positions along the line, $\epsilon=1$ for the first 6 positions, and $\epsilon=-1$ for the other 6 positions, in the blue and red wings of the line, respectively. 
 
\section{Results}

\subsection{Formation of the first penumbral sector and its evolution}

AR NOAA 11490 consists of two preceding main sunspots with positive polarity and a more diffuse negative polarity in the following part of the AR (see Figure 1b). The initial stage of the penumbra formation in the leading spot, located in the FOV A indicated with the dashed box in Figure 1a, was already described in a previous paper \citep{Mur16}. The penumbra forms in about 10 hours and the first sector develops on the side located away from the opposite polarity.

\begin{figure*}[htbp]
\centering
\includegraphics[scale=0.285, clip, trim= 0 200 80 230]{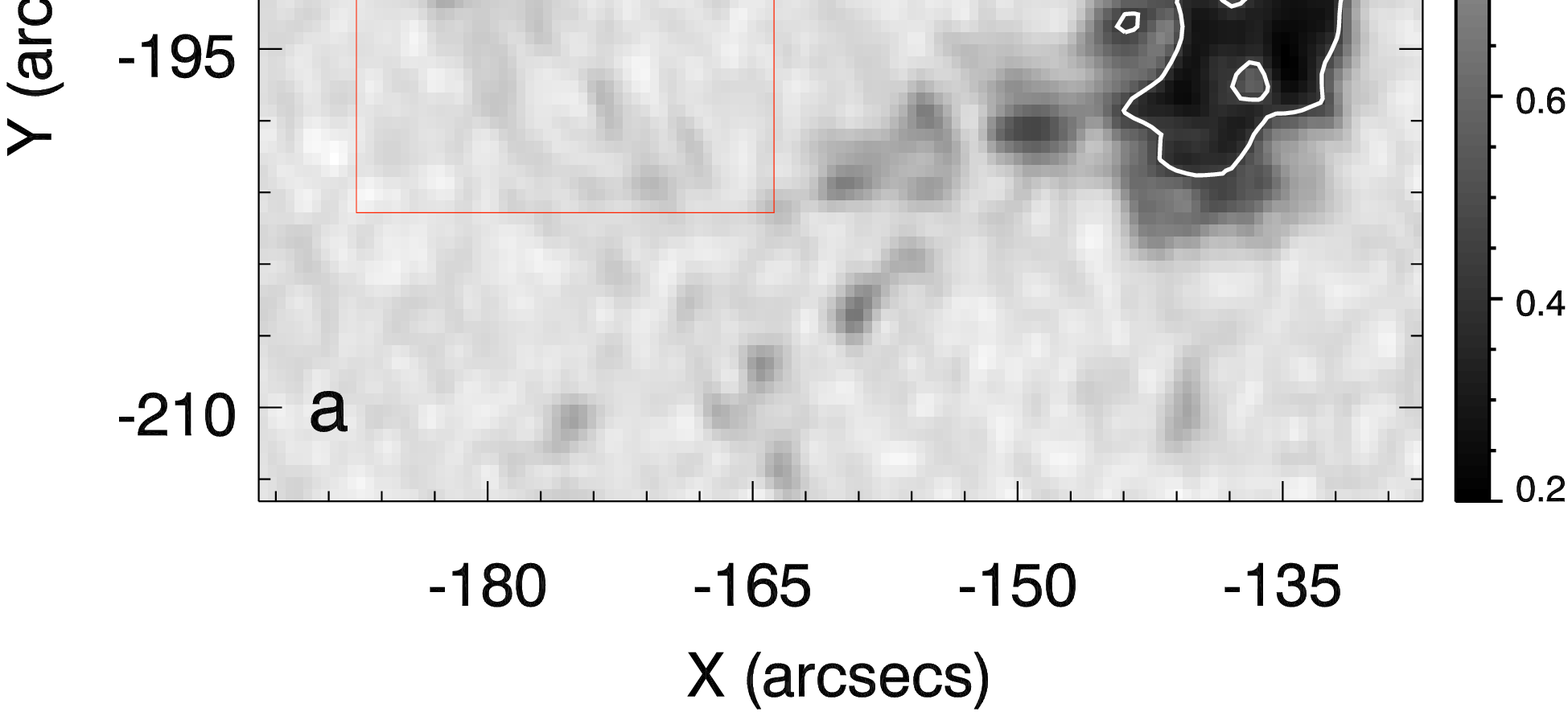}%
\includegraphics[scale=0.285, clip, trim=85 200 80 230]{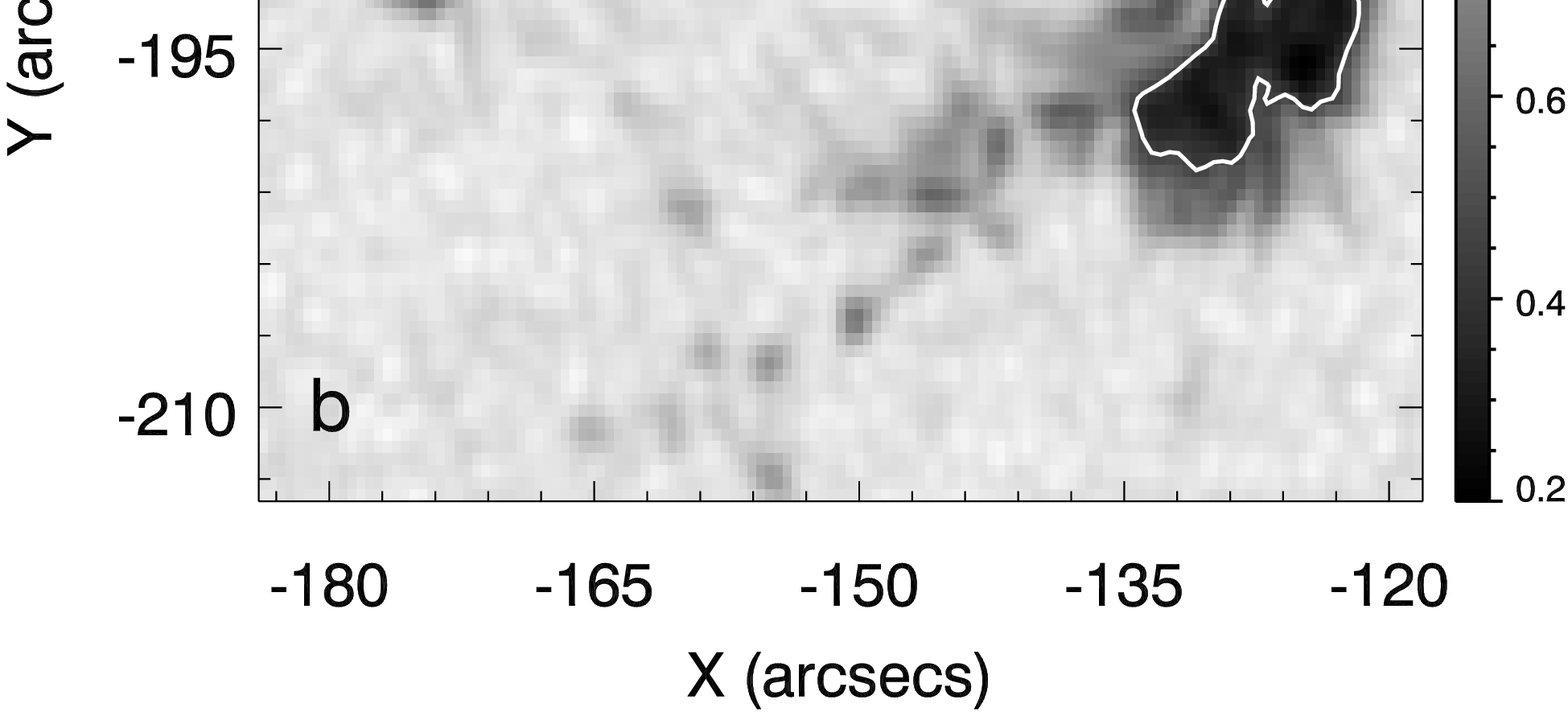}%
\includegraphics[scale=0.285, clip, trim=85 200  0 230]{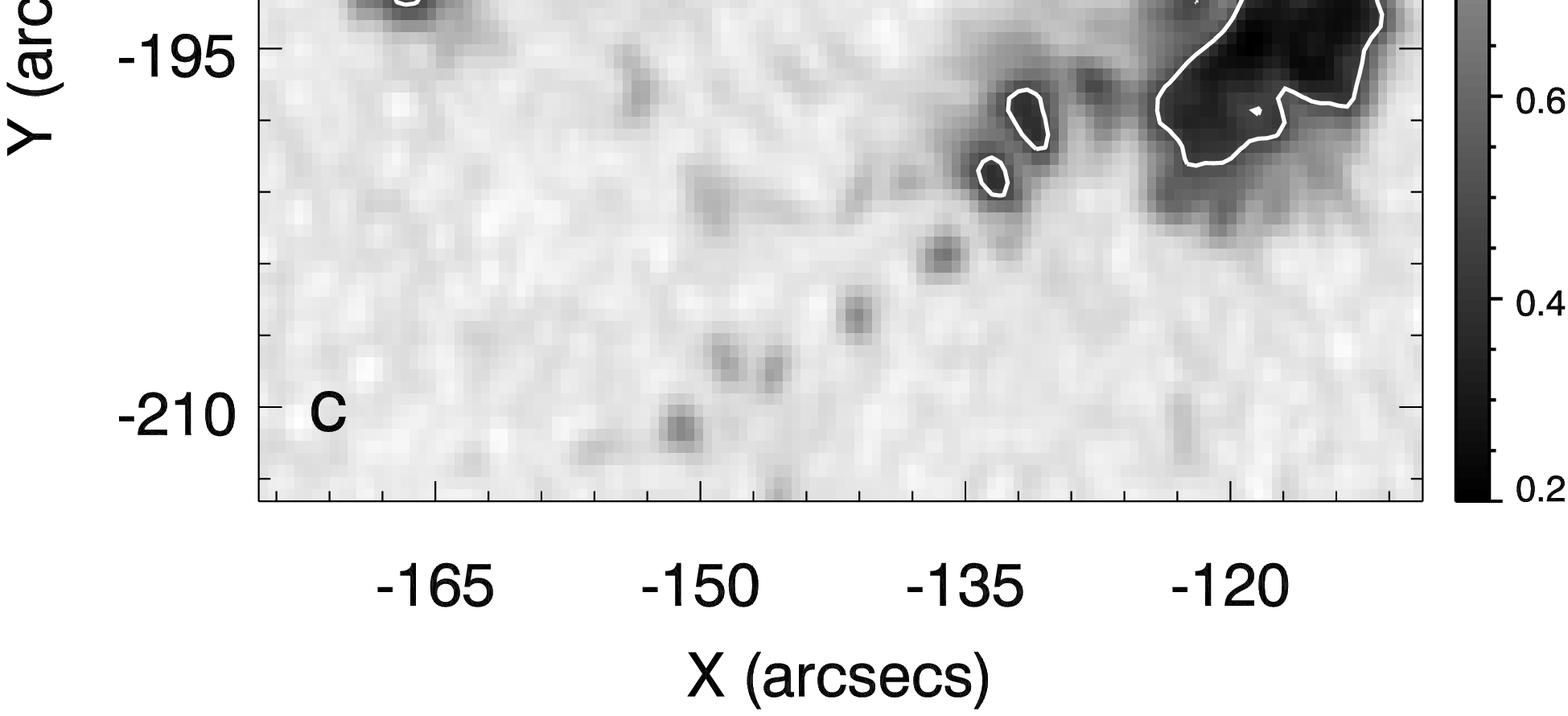}
\includegraphics[scale=0.285, clip, trim= 0 200 80 230]{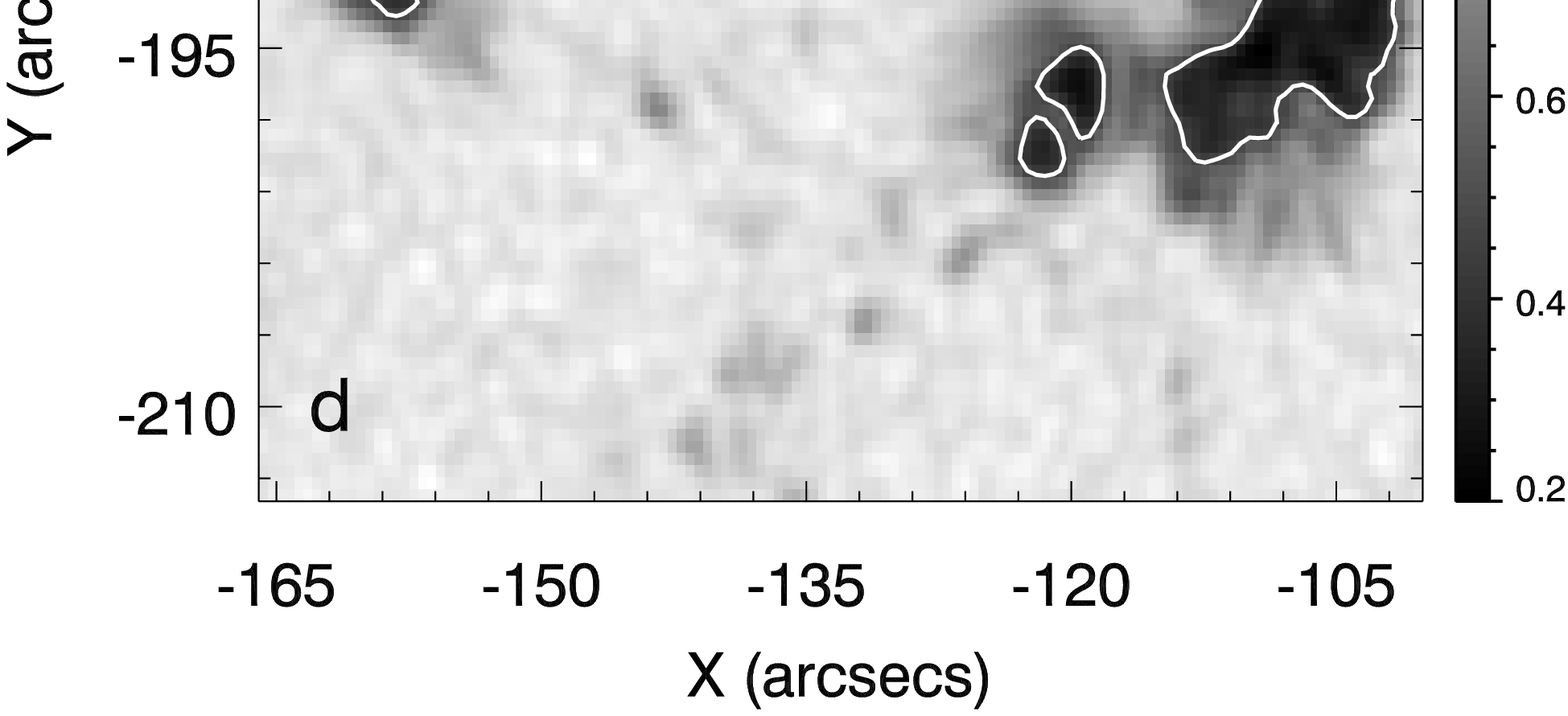}%
\includegraphics[scale=0.285, clip, trim=85 200 80 230]{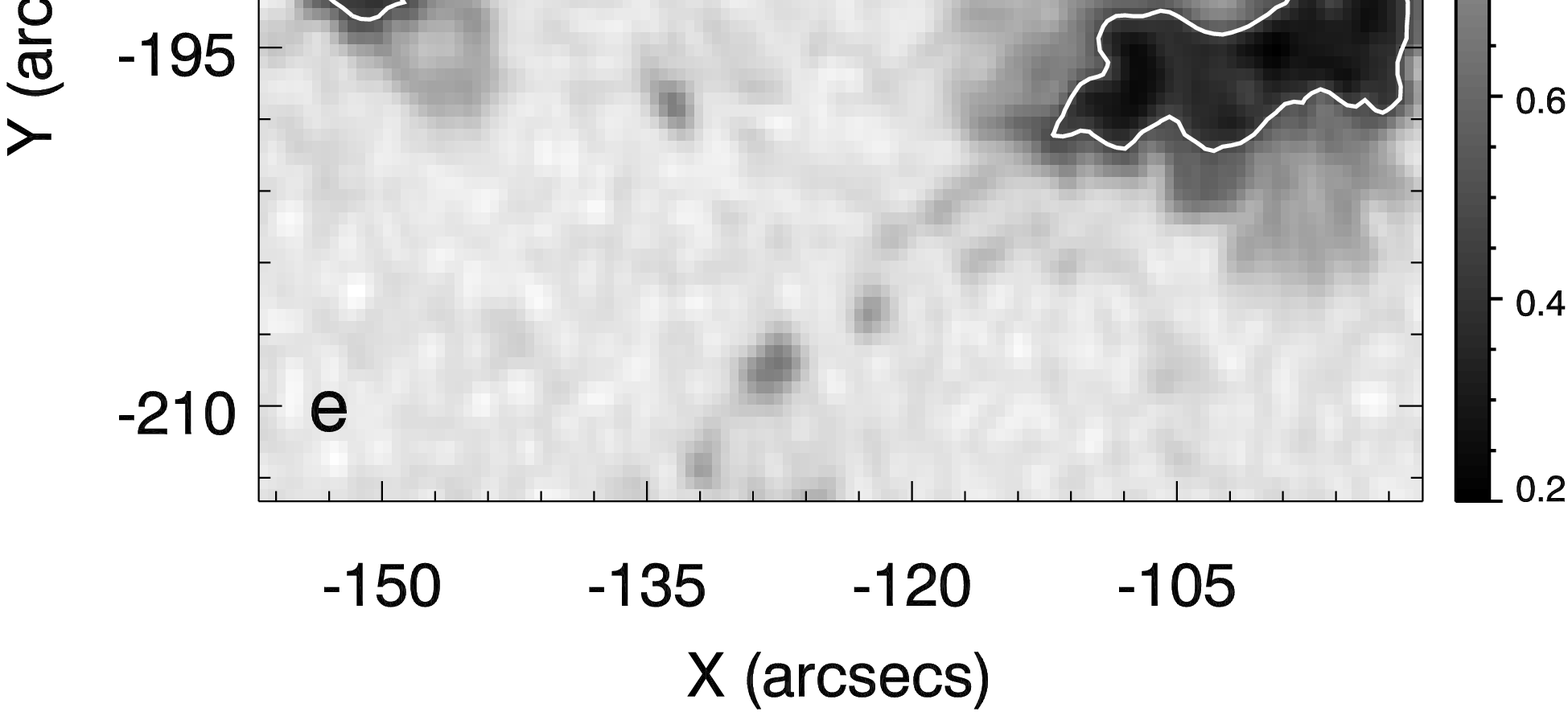}%
\includegraphics[scale=0.285, clip, trim=85 200  0 230]{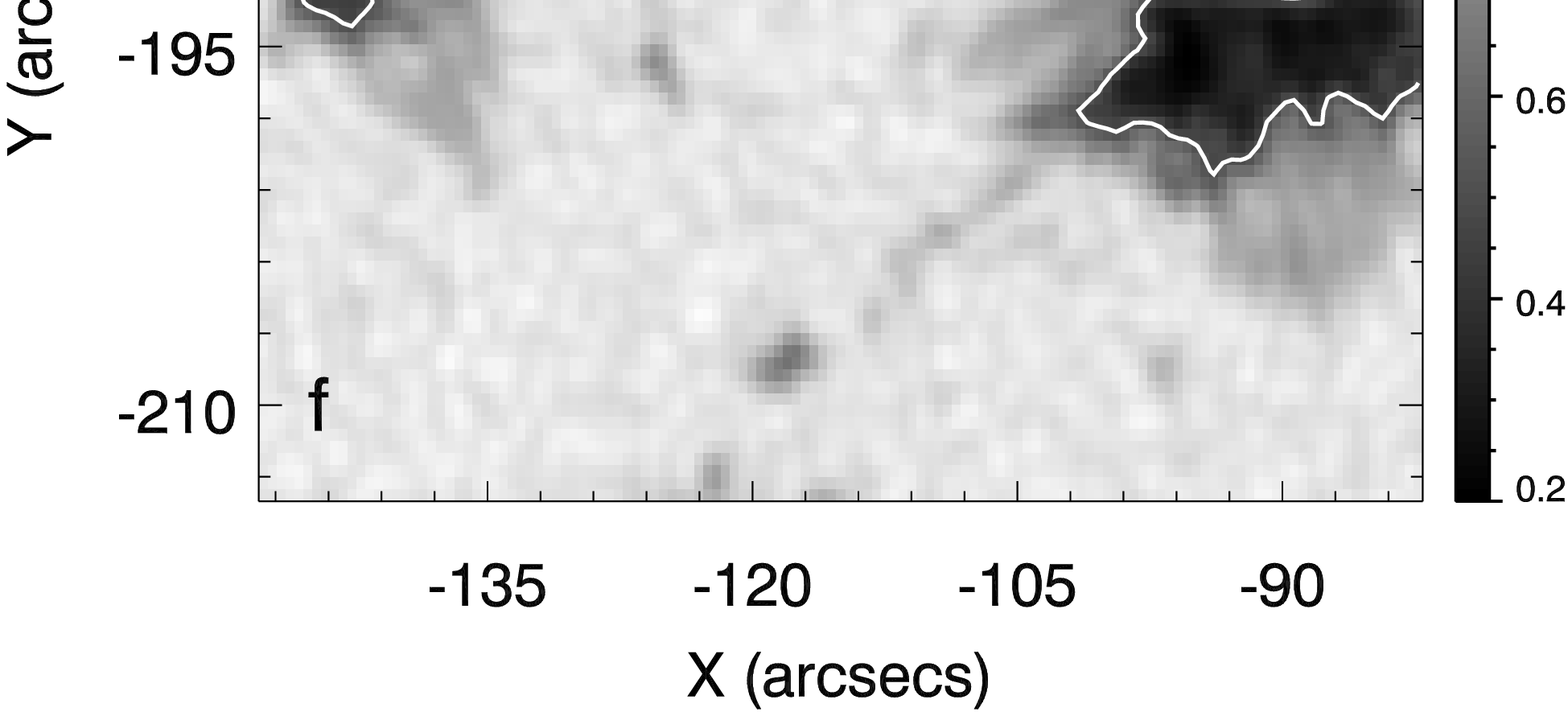}
\includegraphics[scale=0.285, clip, trim= 0 175 80 230]{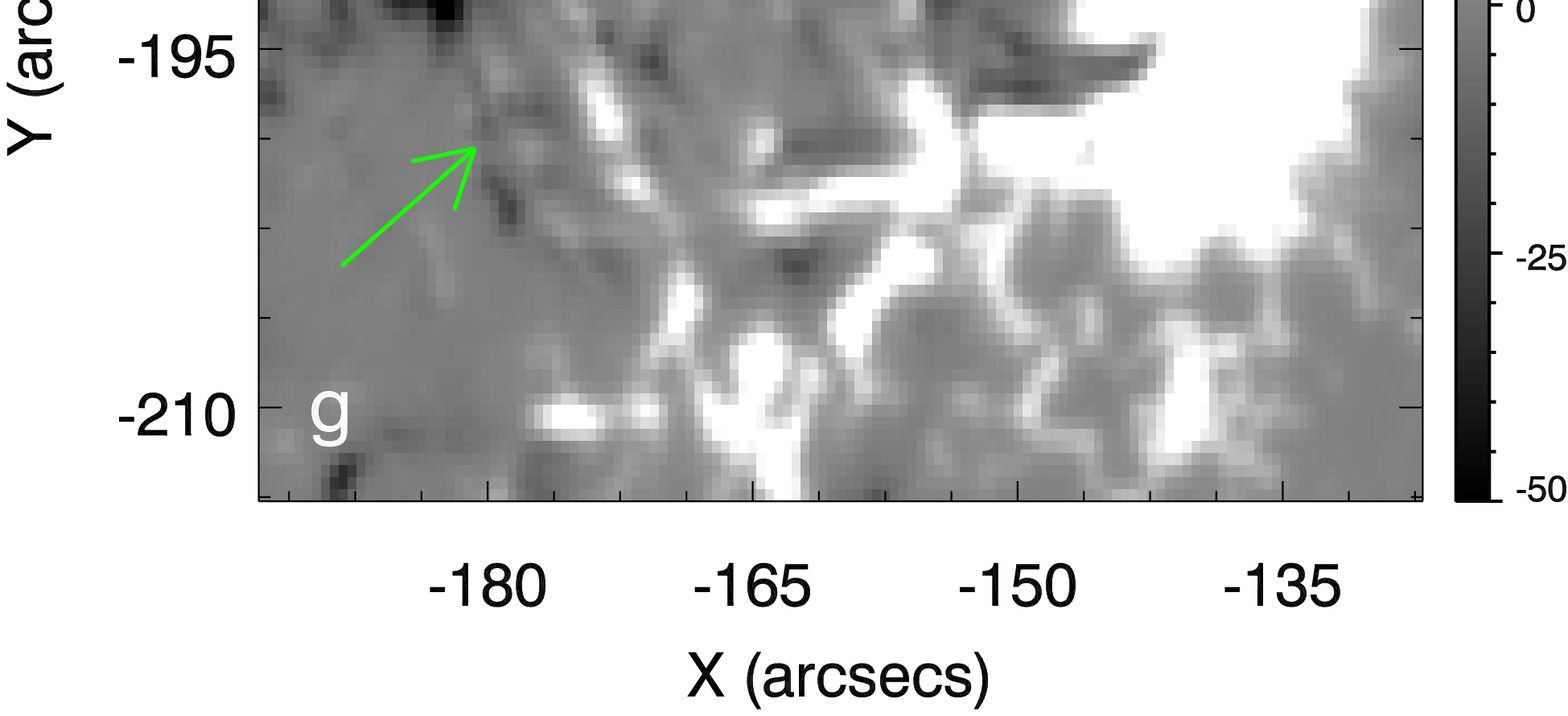}%
\includegraphics[scale=0.285, clip, trim=85 175 80 230]{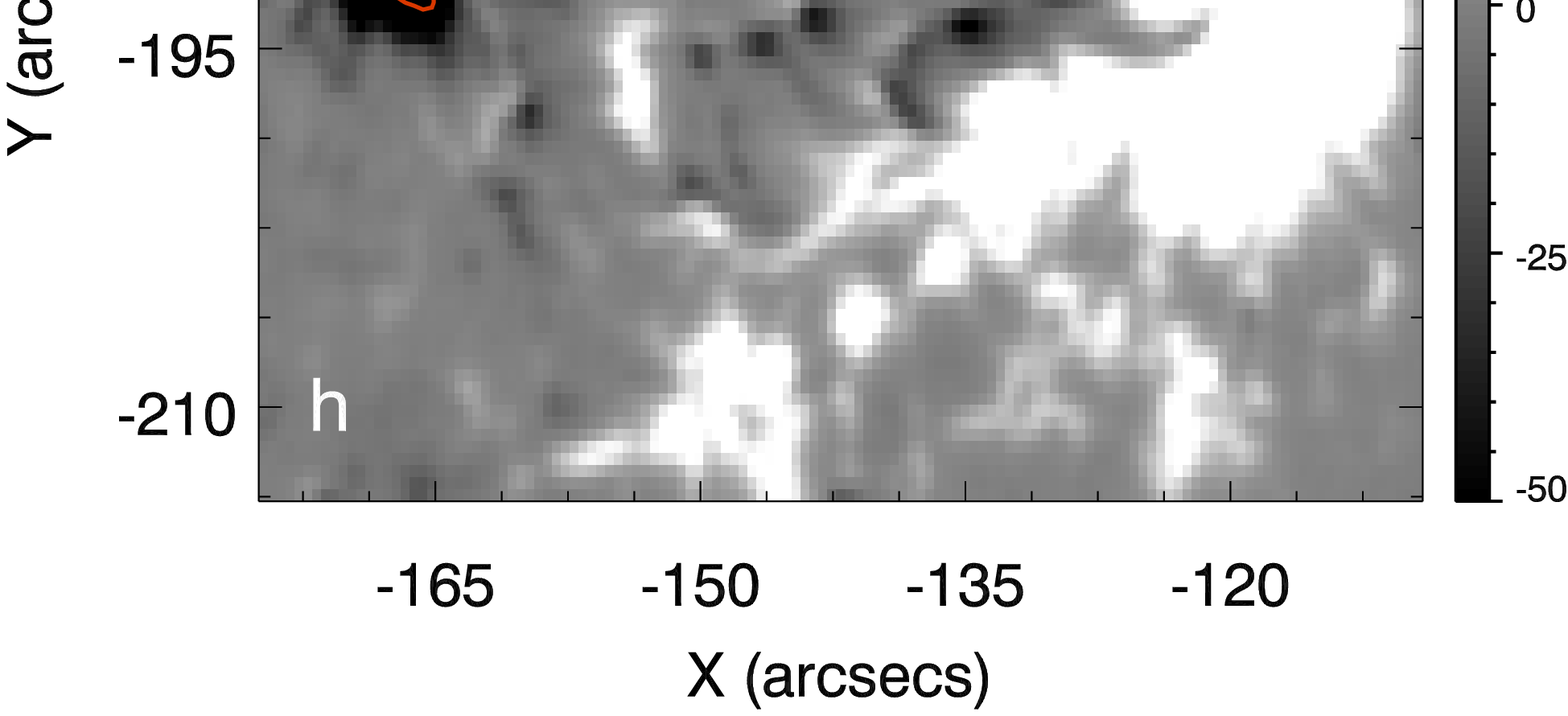}%
\includegraphics[scale=0.285, clip, trim=85 175  0 230]{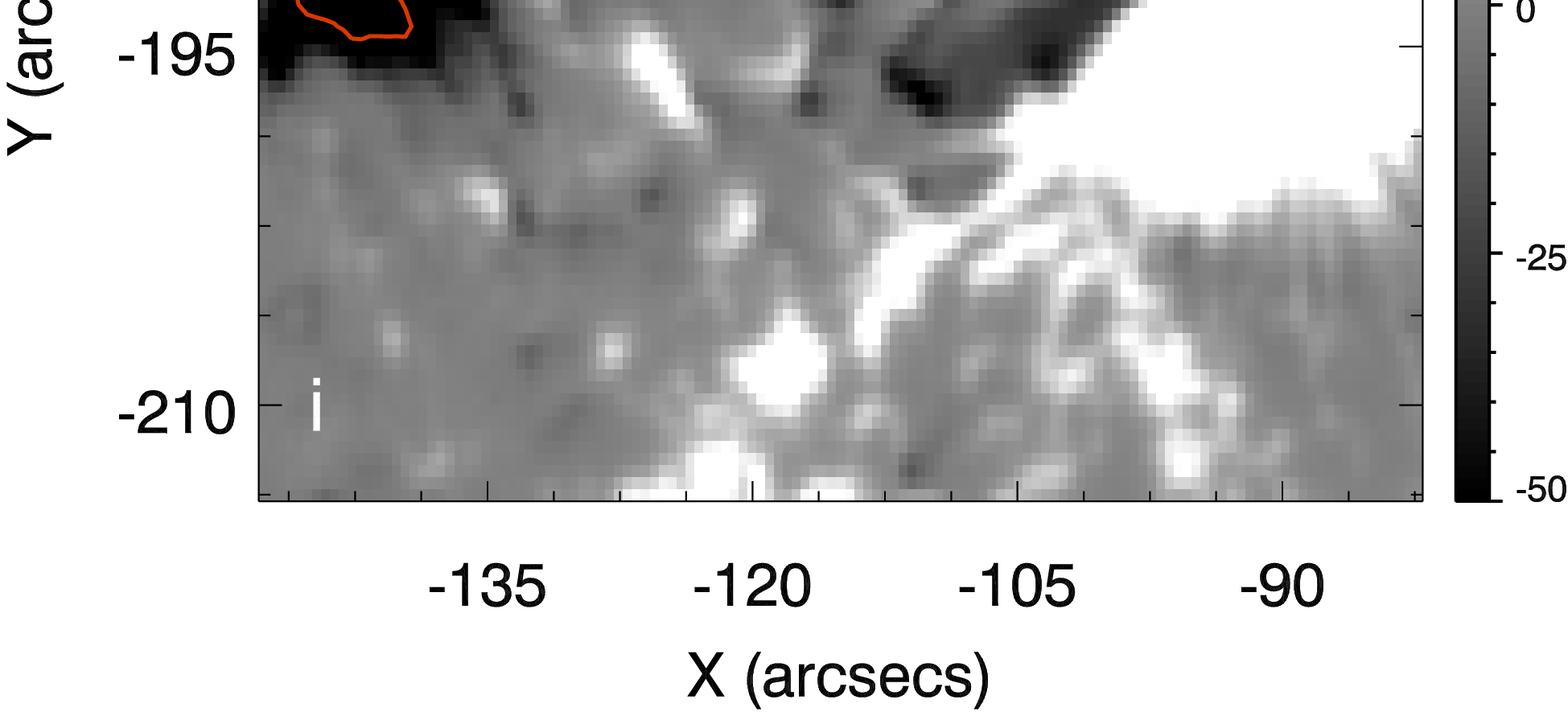}
\caption{(a-f) Continuum filtergrams taken by SDO/HMI on May 28 at the times given at the top of each panel, showing the evolution of AR NOAA 11490. White contours in the continuum images indicate the edge of the pore ($I_{c}=0.5$). The red box in (a) indicates the subFOV displayed in Figura 3. (g-i) LOS magnetograms taken by SDO/HMI on May 28 at the times given at the top left of each panel. The values are saturated at $\pm 500$~G. Red contours in the magnetograms indicate the area characterized by negative magnetic flux density of about $-900$~G.}
\label{fig2}
\end{figure*}

Here, we study the penumbra formation in the following part of the AR, located in the FOV B (solid box, Figure 1a). This phenomenon occurred in about 15 hours. In particular, the pore that is visible at [-195\arcsec,-185\arcsec]$\times$[-205\arcsec,-200\arcsec] evolved into a mature sunspot: a penumbra almost encircled the spot on 29 May 2012 at 21:00 UT (see Figure 1d).

Figure 2 shows the formation of the first penumbral sector of the following negative polarity from 15:00 UT to 20:00 UT on 28 May 2012. We note that from 15:00 UT to 18:00 UT the area of the umbra enlarges as a result of the merging of a patch of the same polarity as the following pore (see the green arrow in Figure 2g). In fact, comparing the images in the continuum of Figure 2a-f we find that the area of the umbra increases of about 42.5 arcsec$^{2}$ in 5 hours (from about 35~arcsec$^2$ at 15:00 UT to 77~arcsec$^2$ at 20:00 UT). The total area of the umbra is determined by contours with a threshold of $I_{c}=0.5$ (see Figure 2a-f). Similarly, the area characterized by negative magnetic flux grows by about 60 arcsec$^{2}$, as derived by contours at $-900$~G (see the magnetograms in Figure 2g-i).

\begin{figure*}[htbp]
	\centering
	\includegraphics[scale=0.385,clip, trim= 70 250 180 340]{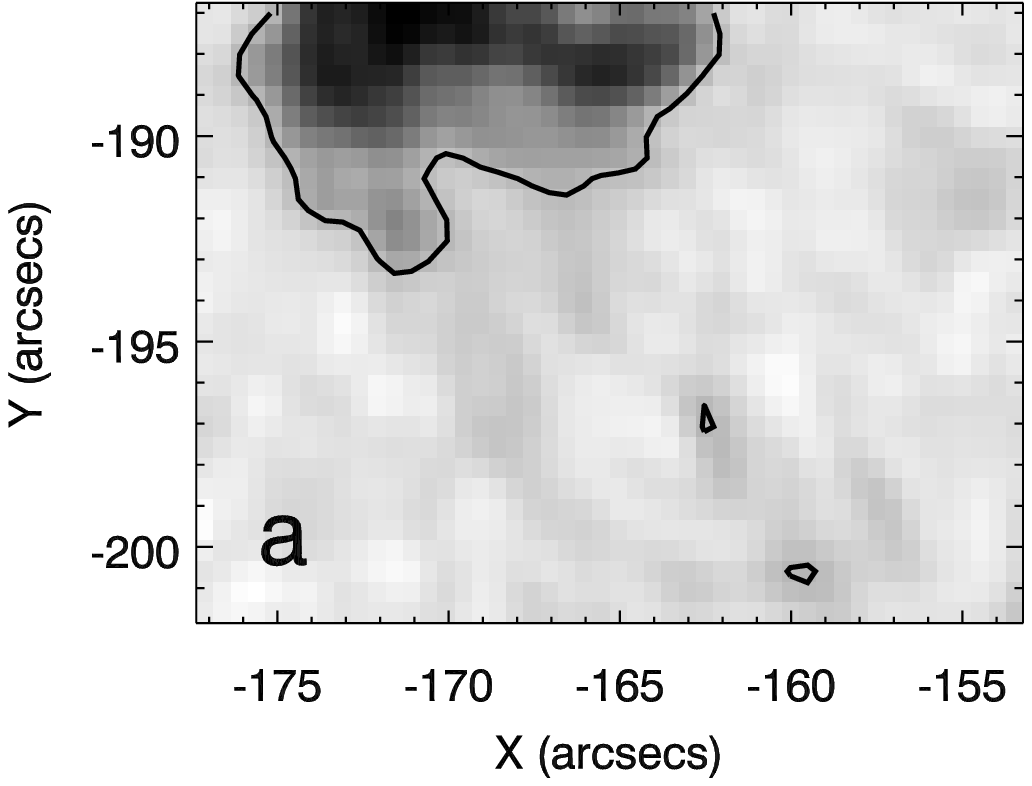}%
	\includegraphics[scale=0.385,clip, trim=115 250 100 340]{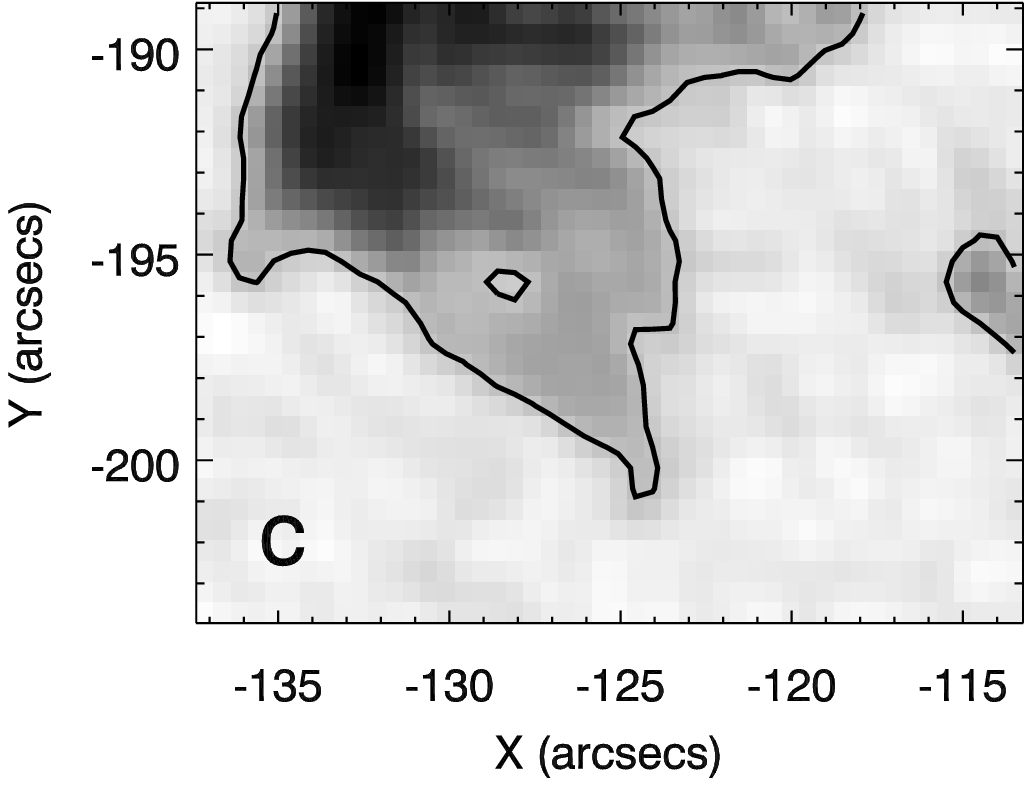}%
	\includegraphics[scale=0.345,clip, trim= 80 140 100 205]{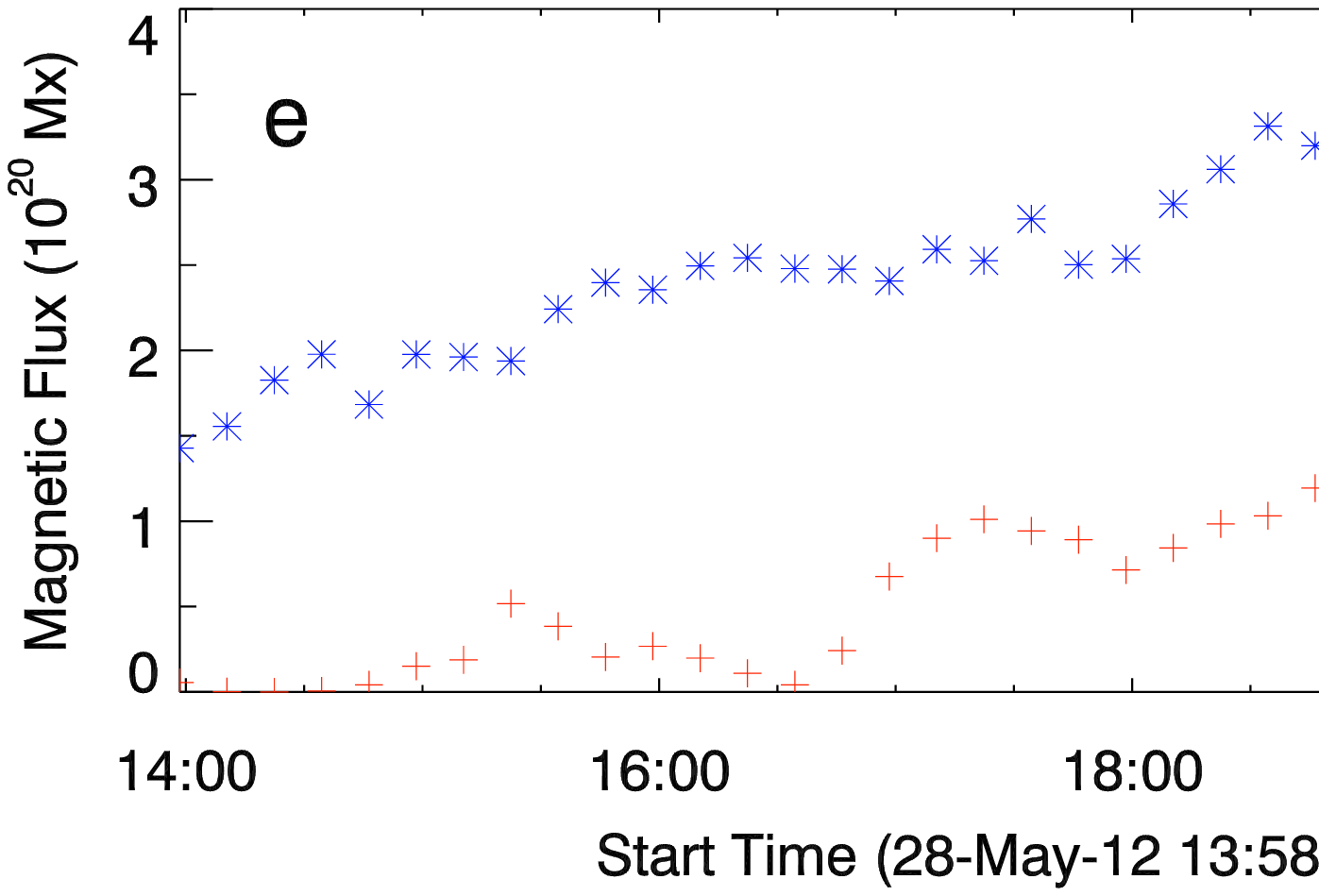}
	\includegraphics[scale=0.385,clip, trim= 70 170 180 340]{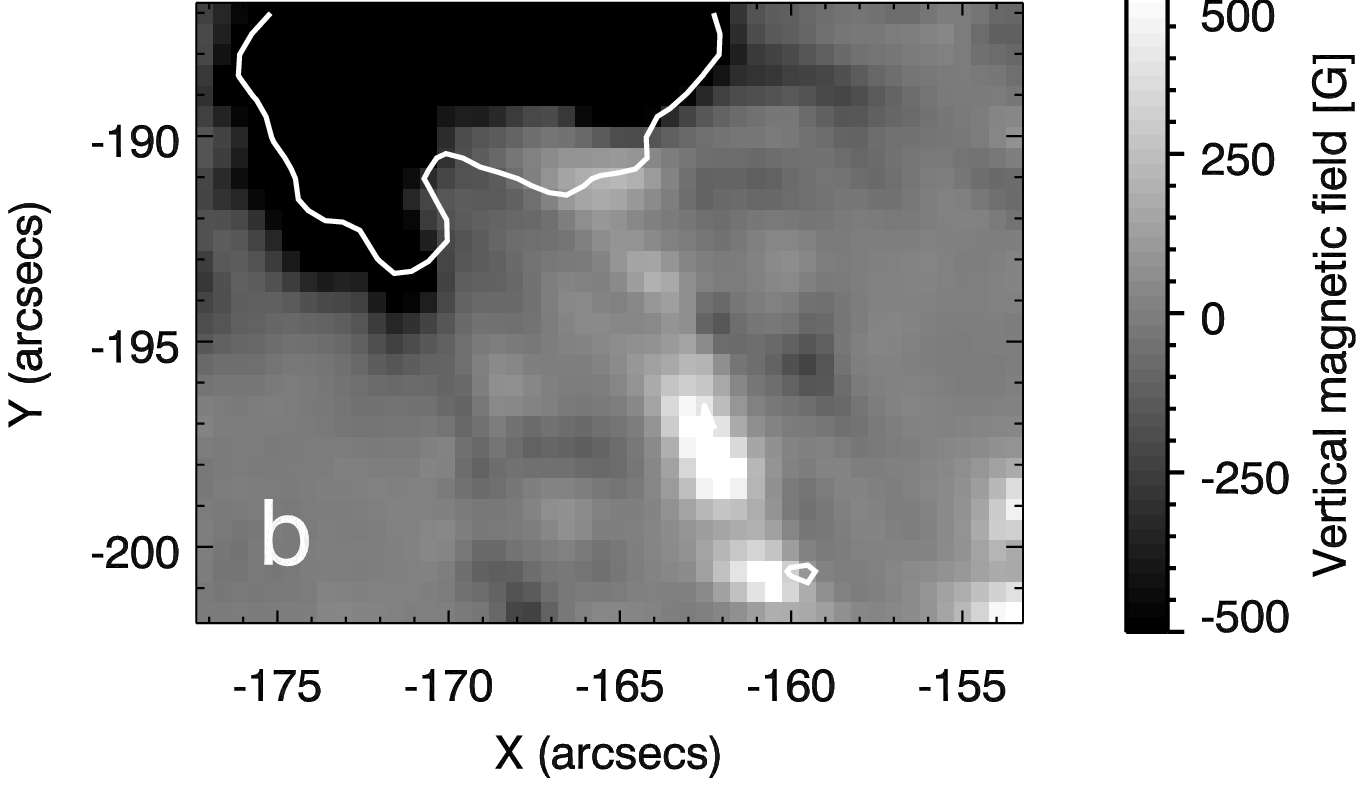}%
	\includegraphics[scale=0.385,clip, trim=115 170 100 340]{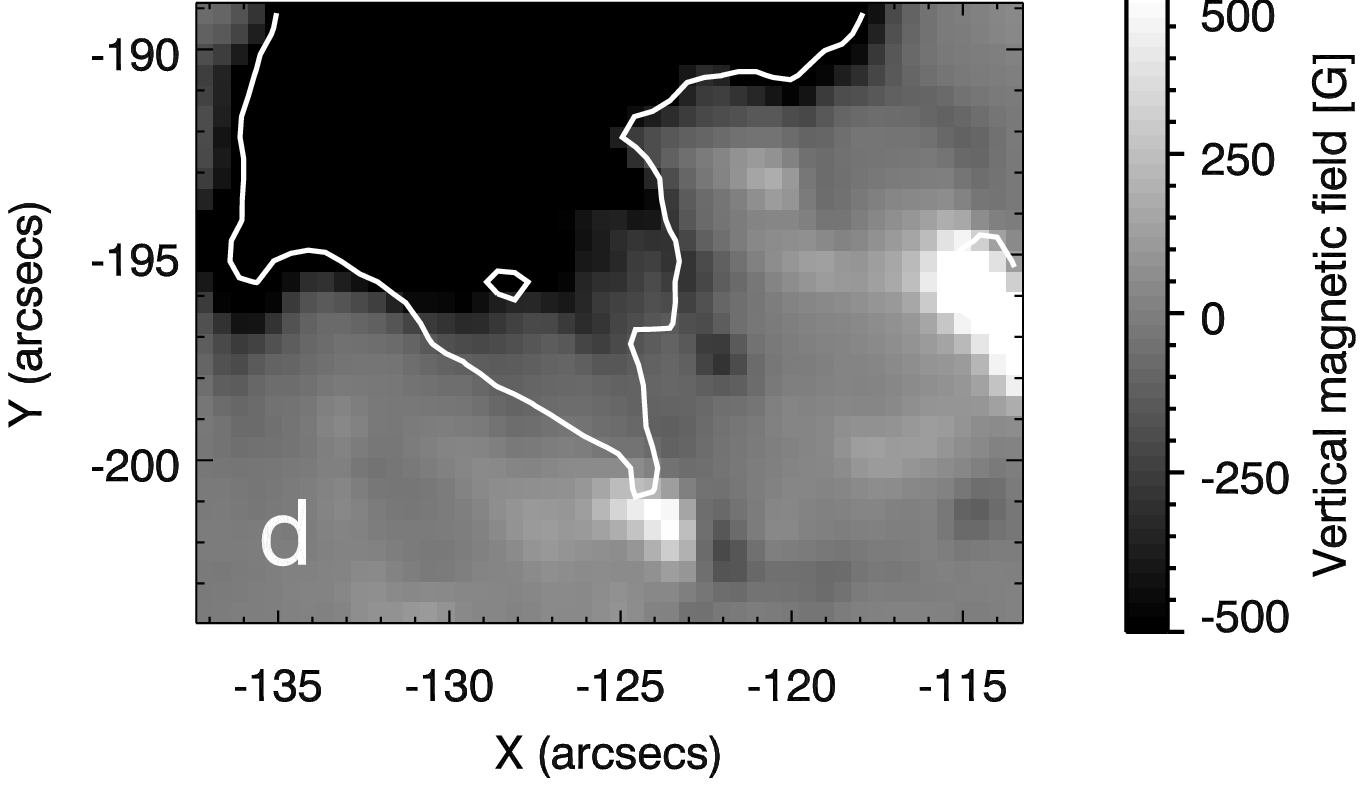}%
	\includegraphics[scale=0.345,clip, trim= 80  85 100 150]{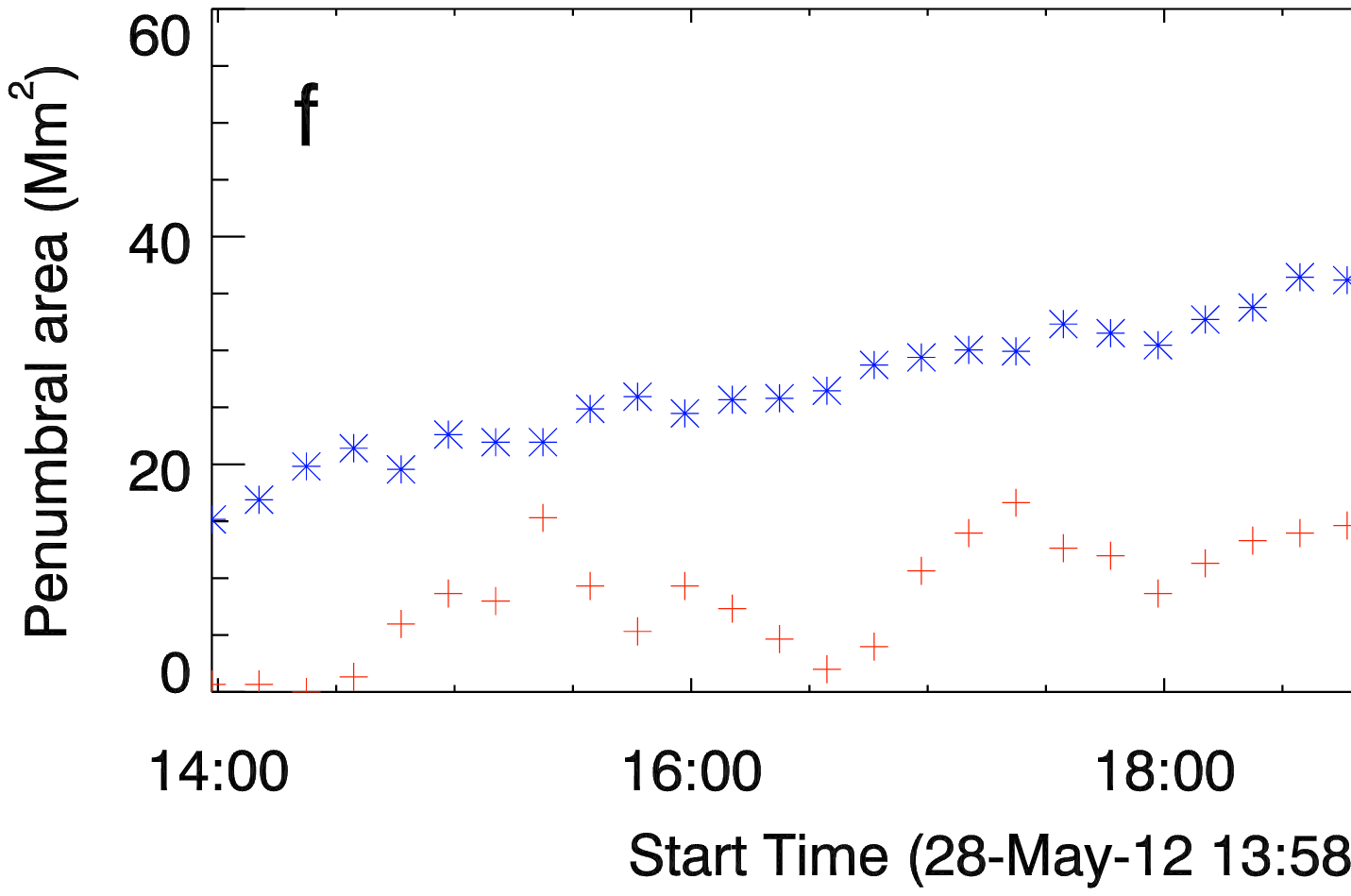}
	\caption{Left: Maps of the continuum intensity and of the vertical component of the magnetic field, at 15:00 UT (3a and 3b) and at 20:00 UT (3c and 3d), respectively. Values of the vertical magnetic field are saturated at $\pm 500$~G. Black (white) contours indicate the area characterized by $I_{c}=0.9$. Right: Plots of the magnetic flux within the penumbral area (e) and of the total penumbral area (f) comprised in the box. Red (blue) symbols refer to positive (negative) areas. Positive values are amplified by a factor of 5 for a better visualization.}
	\label{fig3}
\end{figure*}

At 18:00 UT we can see that the first penumbral filaments are formed. In the subsequent three hours they increase radially in size, outwards from the spot and toward the opposite polarity. Figure 3 clarifies that these penumbral filaments are stable. We analyze more in detail the region around the first penumbral sector, indicated with the box in Figure 2a, from 14:00 UT to 21:00 UT. To get rid of possible projection errors, we use the vertical component of the magnetic field deduced from SHARP data. In Figure 3a, at 15:00 UT we find the presence of some dark areas with $0.5 < I_{c} < 0.9$. The larger area at [-175\arcsec,-165\arcsec]$\times$[-193\arcsec,-188\arcsec] is characterized by negative polarity, the same as the adjacent pore. There is also a dark patch with positive field facing the pore at [-164\arcsec,-159\arcsec]$\times$[-201\arcsec,-196\arcsec] (see Figure 3b). However, at 20:00 UT the latter disappears, while the dark area with negative field adjacent to the pore increases, as can be seen from Figure 3c-d. In Figure 3e we plot the positive (red symbols) and negative (blue symbols) magnetic flux evolution in the areas with $0.5 < I_{c} < 0.9$, corresponding to the regions of penumbra formation comprised in the box. The simultaneous evolution of the spatial extent of these areas is also reported in Figure 3f. These graphs illustrate that the first penumbral sector grows smoothly without interruption as the negative magnetic flux increases. This trend remains constant in the following hours after 21:00 UT. By contrast, the penumbral areas characterized by positive magnetic field are transient. 

\subsection{Elongated granules and Arch Filament System}

Figure 4 shows the whole region observed with IBIS, that is both FOV A and FOV B, observed from 13:39 UT to 14:12 UT and from 14:19 to 14:38 UT, respectively. In particular, in this Figure (top panels) we display the maps acquired in the continuum of the \ion{Fe}{1} 617.3 nm and in the center of the \ion{Ca}{2} 854.2 nm line on May 28. To obtain these images, where both the following and the leading polarity observed by IBIS are visible, we used a cross-correlation technique aligning a common area between the two FOVs. 
 
The map of the continuum (top left panel, Figure 4) reveals that the following spot is characterized by an umbra and a region between the two polarities comprised in [-170\arcsec,-150\arcsec]$\times$[-225\arcsec,-205\arcsec], where elongated granules are visible.
Figure 4 (top right panel) shows the same region in the \ion{Ca}{2} line center. We note the presence of an AFS. This suggests that the area characterized by elongated granules in photosphere corresponds to a region of magnetic flux emergence. This region is also characterized by a magnetic field configuration having a filamentary shape (Figure 4, bottom left panel) with a sea-serpent configuration, i.e., formed by a bundle of bipolar patches \citep[e.g.][]{SainBel08}. In the magnetic field map (Figure 3, bottom left panel) we note that these filamentary structures are not aligned with the structure of the AFS visible in the \ion{Ca}{2} line center. The filamentary and mixed pattern of the magnetic field can be also distinguished in the inclination angle map (see Figure 4, bottom right panel), but on a larger scale. In fact, we find filaments characterized in their inner part by magnetic field inclination both less and larger than 90$^{\circ}$.

\begin{figure*}[htbp]
	\centering
	\includegraphics[scale=0.325,clip, trim=0   70 10 110]{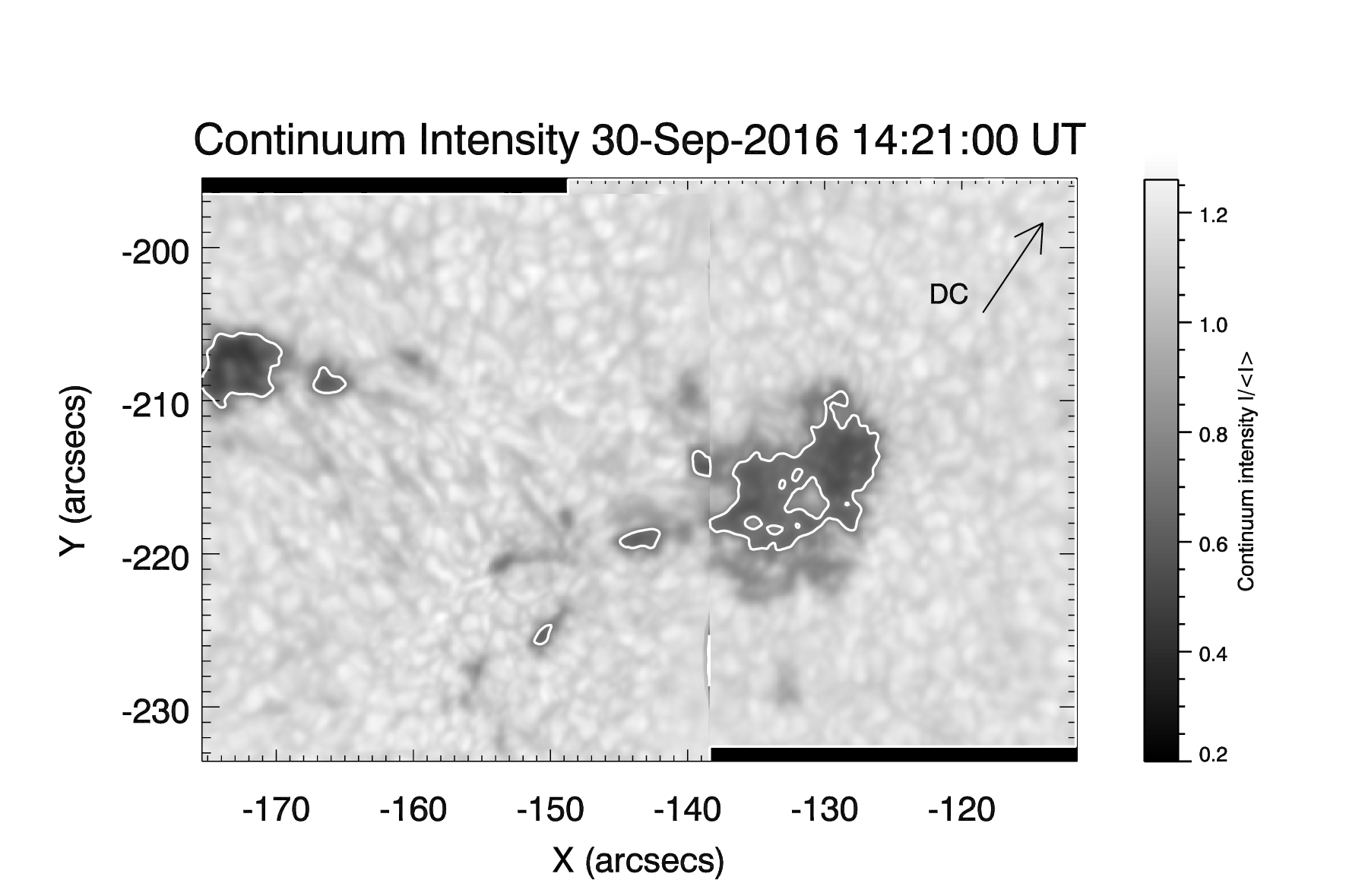}%
	\includegraphics[scale=0.325,clip, trim=125 70 30 110]{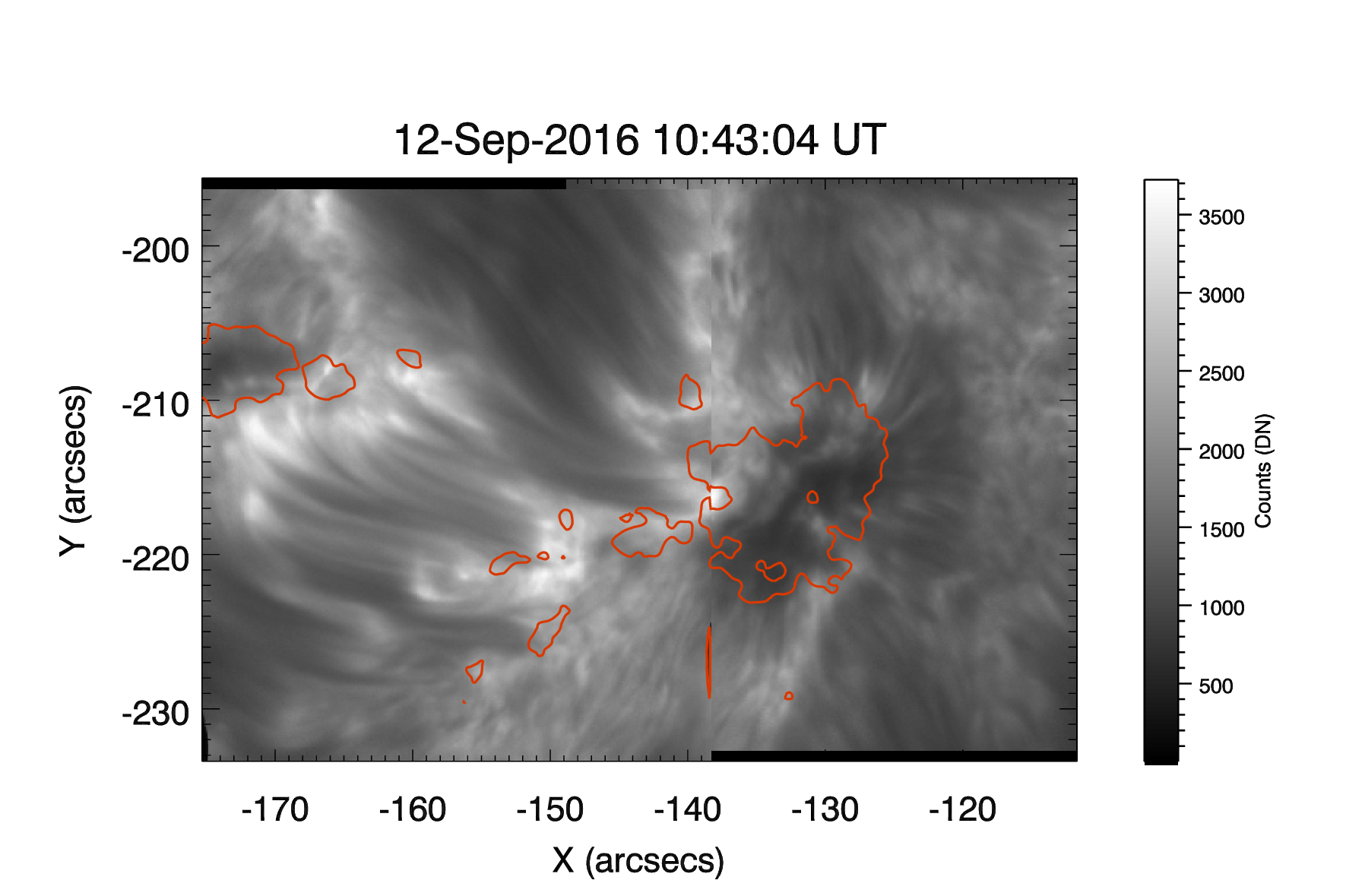}
	\includegraphics[scale=0.325,clip, trim=0    0 10 110]{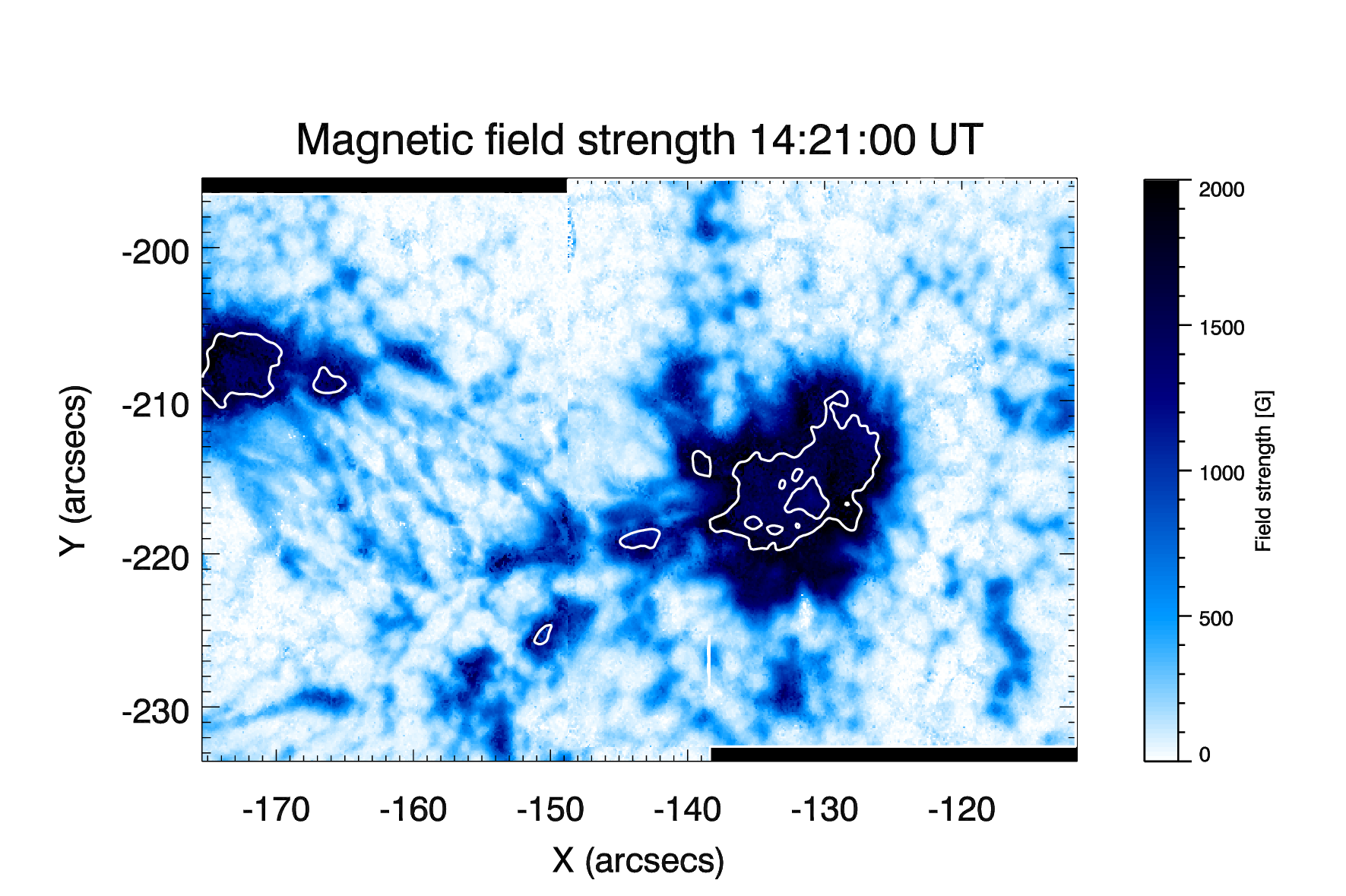}%
	\includegraphics[scale=0.325,clip, trim=125  0 30 110]{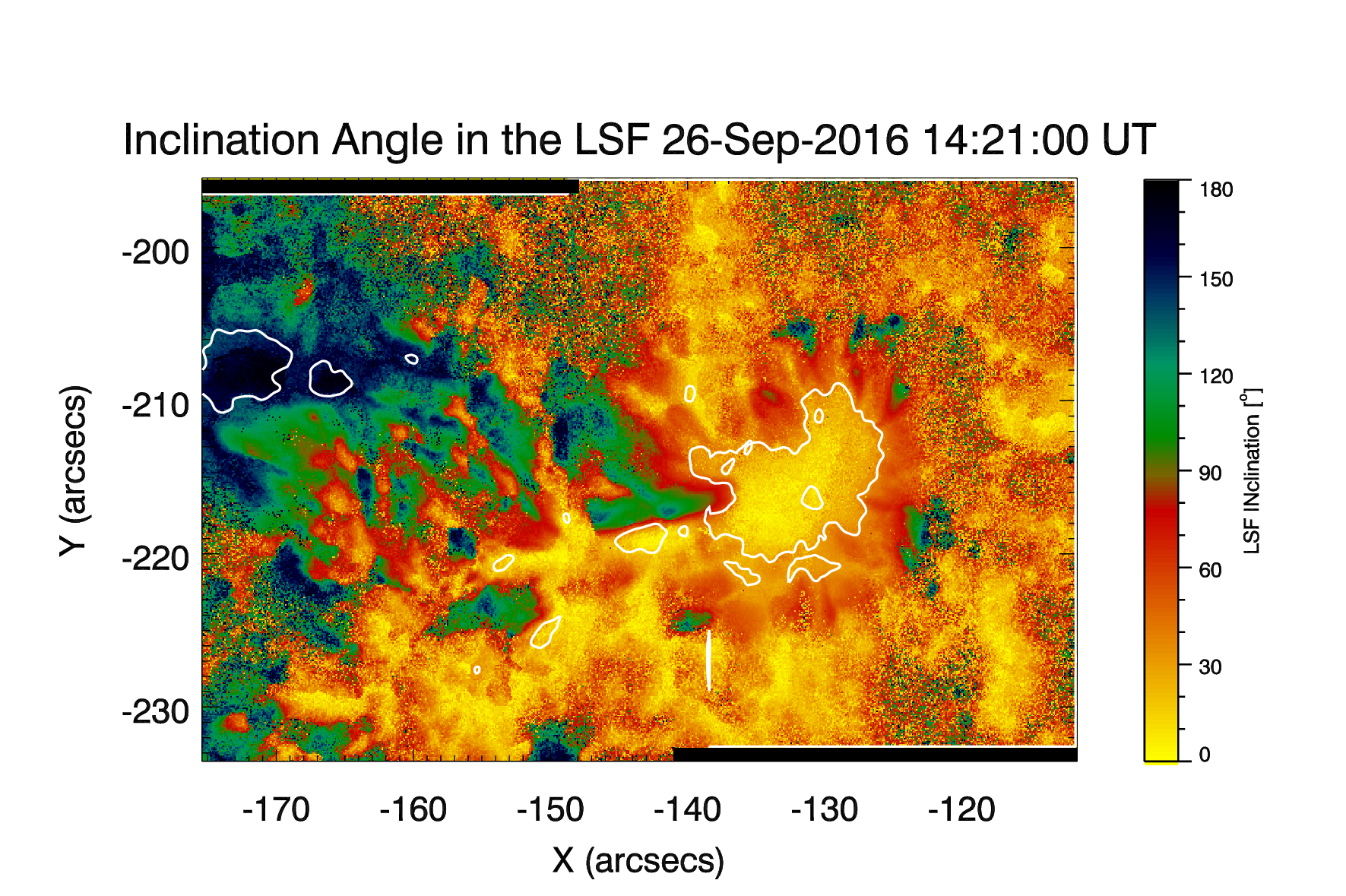}
	\caption{Top panels: maps of continuum intensity at \ion{Fe}{1} 617.3 nm line and in the center of the \ion{Ca}{2} 854.2 nm line on 2012 May 28 at around 14:20 UT from IBIS data. Bottom panels: maps of the magnetic field strength and of the inclination angle obtained from the SIR inversion of the Stokes profiles along the \ion{Fe}{1} 617.3 nm line acquired by IBIS. The contours indicate the edge of the pore at $I_{c} = 0.8 $.}
	\label{fig4}
\end{figure*} 

\begin{figure*}[htbp]
	\centering
	\includegraphics[scale=0.50,clip, trim=0  180 60 240]{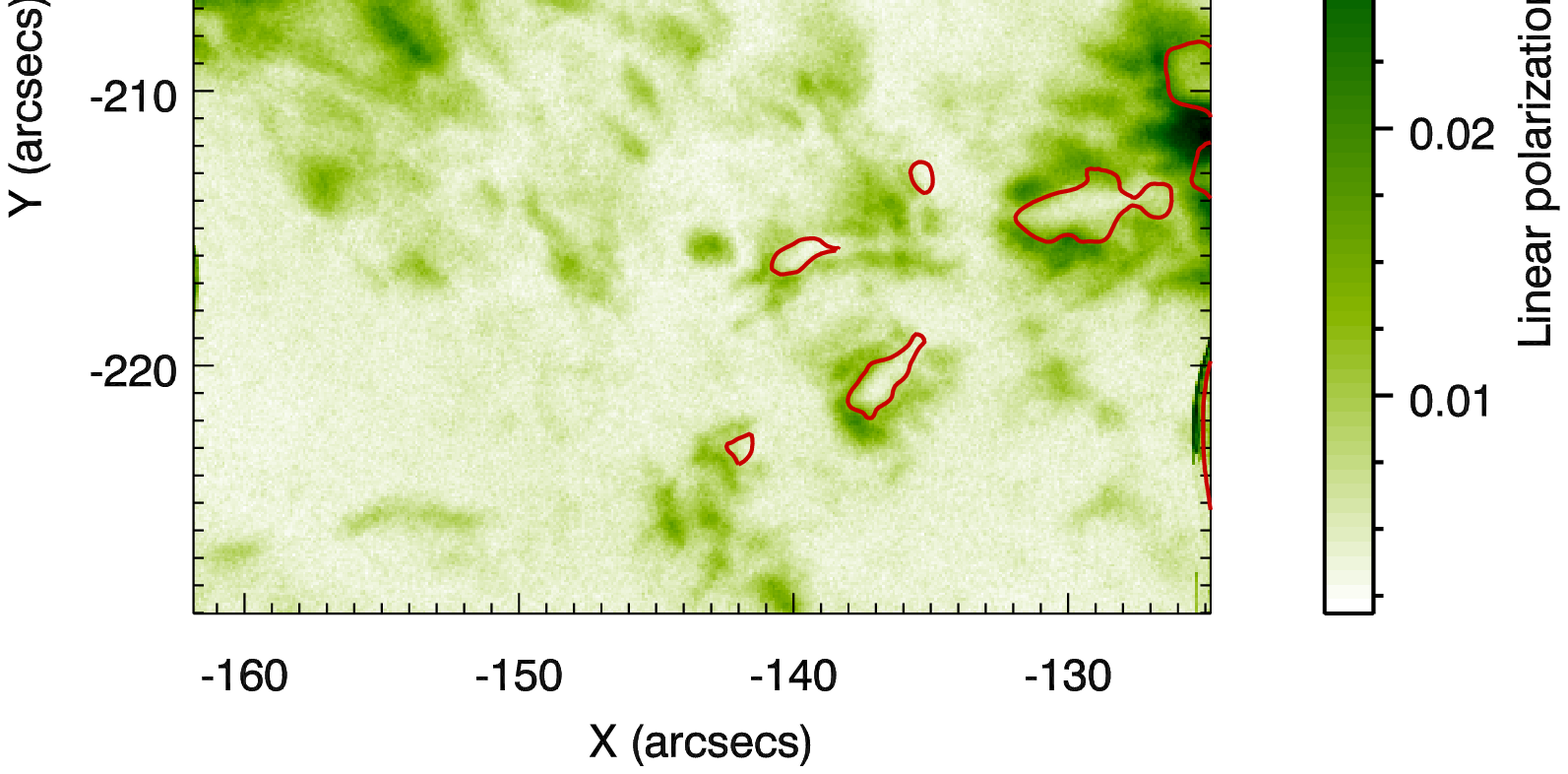}%
	\includegraphics[scale=0.50,clip, trim=84 180 90 240]{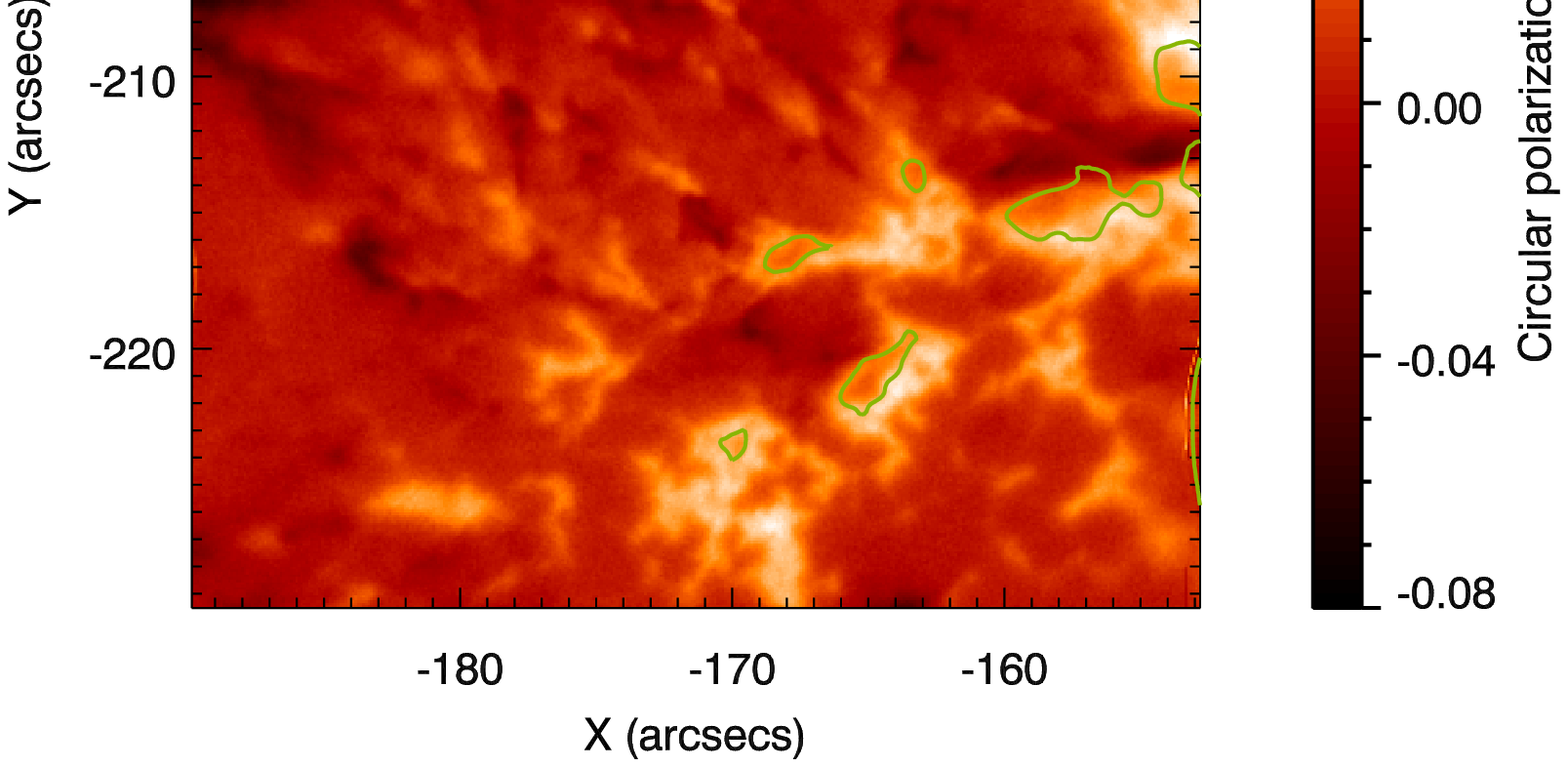}
	\caption{Maps of mean linear and circular polarization signals, $L_{s}$ (left) and $V_{s}$ (right), obtained from IBIS data for the FOV B. The red and green contours indicate the edge of the pore at $I_{c} = 0.8 $ as seen in the simultaneous continuum intensity image (see Figure 4).}
	\label{fig5}
\end{figure*}

This filamentary structure can be also recognized in the linear and circular polarization maps relevant to the FOV B, which are displayed in Figure 5. These maps are deduced by IBIS spectropolarimetric data at 14:21 UT, i.e., the scan with the best seeing. We can use these maps as proxies of vertical and horizontal fields, respectively, given the position close to the disc center of the AR.

In the $L_{s}$ map (Figure 5, left panel), we note around the proto-spot a wide region at [-210\arcsec,-205\arcsec]$\times$[-162\arcsec,-152\arcsec] with different values of linear polarization. In particular, in the region where later the first penumbral sector starts to form, the values are larger than 4$\%$, while in the northern part of the pore the linear polarization is smaller than about 1$\%$. The $V_{s}$ map (Figure 5, right panel) shows a diffuse negative polarity corresponding with the area of the proto-spot and its surroundings. The area between the two magnetic polarities is characterized by a pattern of alternate positive and negative bipoles, with a sea-serpent configuration.

\section{Discussion and Conclusions}
Our findings concerning the penumbra formation in the following polarity of the AR NOAA 11490 are that: (i) penumbral filaments form in the side facing the opposite polarity; (ii) elongated granules in the photosphere, an AFS in the chromosphere, and a magnetic field with sea-serpent configuration are present in the region between the two polarities.

The region between the two polarities is characterized by elongated granules, as described in previous observations (see \citealp{Sch10a}). However, \citet{Sch10a} concluded that this region, being characterized by flux emergence, cannot form stable penumbrae. Nevertheless, our observations highlight that the first stable penumbral sector around the pore forms exactly in the region where a large amount of magnetic flux is emerging, as demonstrated by the presence of the AFS in the chromosphere. 

The filamentary pattern of the magnetic field, seen also in the inclination angle of the magnetic field, in the polarization map, and in the continuum, is similar to the sea-serpent configuration. Bipolar patches are observed close to the proto-spot. These features are interpreted as being produced by field lines arising (positive polarity) and going down (negative polarity) in the photosphere. Spectropolarimetric measurements have revealed the existence of such field lines that originate in the mid penumbra and propagate toward the outer penumbral boundary \citep[see, e.g.][]{SainBel08,Kitiashvili10}. \citet{Weiss04} suggested that the turbulent pumping by granular convection drags the flux tubes downward in the moat region to form the sea-serpent configuration. This magnetic configuration, due to magnetoconvection in strongly inclined magnetic fields \citep{Kitiashvili10}, is frequently adopted also to interpret observed properties of bipolar magnetic features in relation with penumbral structures \citep{SainPil05,SainBel08}. In this sense, this configuration may be considered as a further precursor of the penumbra formation, in cases similar to our observations.

\citet{Lim13} observed the formation of penumbral filaments associated with flux emergence under pre-existing chromospheric canopy fields. In particular, they linked the presence/absence of a pre-existing horizontal field in the chromosphere not only to the penumbra formation, but also to another issue related to the penumbra, such as the formation of an orphan penumbra.

Recently, \citet{MacTagg} presented simulations of AR flux emergence where they demonstrated that the overlying magnetic canopies are a consequence of how the magnetic field emerges into the photosphere. These canopies mainly form in the region away from the opposite polarities, where the penumbra first forms. However, also in the region between the two main sunspots of the AR there is a low probability, but not null, of finding near-horizontal fields. \citet{MacTagg} suggest that such canopies are the source of highly inclined horizontal fields required to produce penumbrae. This situation may explain the presence of horizontal fields at the photospheric level before the penumbra formation found in our observations (Figure 4, left panel). 

From our study we think that formation of the first penumbral sector between the two polarities may be ascribed not only to the presence of strong overlying canopy filds, as suggested by \citet{Lim13}, but also to the rising velocity of the rising velocity of the magnetic flux bundle, which might differ from the other cases presented in the literature. Therefore, it would be important to collect new data sets relevant to penumbra formation, hopefully occurring between the two main polarities of an AR, to provide observational constraints for the rising velocities. This will also allow to have a larger sample of this kind of evolution, in order to unveil the underlying physical mechanism.

\acknowledgments

The authors are grateful to an anonymus referee for constructive comments. The authors wish to thank the DST staff for its support during the observing campaigns. The authors also thank G. Cauzzi for useful information. The research leading to these results has received funding from the European Commission's Seventh Framework Programme under the grant agreement SOLARNET (project no. 312495). This work was also supported by the Istituto Nazionale di Astrofisica (PRIN-INAF-2014), by the University of Catania (PRIN-MIUR-2012) and by Space Weather Italian COmmunity (SWICO) Research Program.



{\it Facilities:} \facility{DST (IBIS)}, \facility{SDO (HMI)}





\clearpage


\begin{thebibliography}

\bibitem[Bobra et al.(2014)]{Bobra14} Bobra, M. J., Sun, X., Hoeksema, J. T., et al.\ 2014, \solphys, 289, 3549

\bibitem[Bruzek(1980)]{Bruz80} Bruzek, A.\ 1980, Solar and Interplanetary Dynamics, 91, 203 

\bibitem[Cavallini(2006)]{Cav06} Cavallini, F. 2006, Sol. Phys., 236, 415

\bibitem[Cheung et al.(2007)]{Cheung07} Cheung, M.~C.~M., Sch{\"u}ssler, M., \& Moreno-Insertis, F.\ 2007, \aap, 467, 703

\bibitem[Cheung et al.(2008)]{Cheung08} Cheung, M.~C.~M., Sch{\"u}ssler, M., Tarbell, T.~D., \& Title, A.~M.\ 2008, \apj, 687, 1373

\bibitem[Georgoulis (2005)]{Geo05}  Georgoulis, M. K. 2005, \apj, 629, L69

\bibitem[Kitiashvili et al.(2010)]{Kitiashvili10} Kitiashvili, I.~N., Bellot Rubio, L.~R., Kosovichev, A.~G., et al.\ 2010, \apjl, 716, L181

\bibitem[Leka \& Skumanich(1998)]{Lek98} Leka, K.D., \& Skumanich, A. 1998, \apj, 507, 454

\bibitem[Lim et al.(2013)]{Lim13} Lim, E.-K., Yurchyshyn, V., Goode, P., \& Cho, K.-S.\ 2013, \apjl, 769, L18 

\bibitem[L\"ofdahl(2002)]{Lof02} L\"ofdahl, M.G. 2002, SPIE, 4792, 146L

\bibitem[MacTaggart et al.(2016)]{MacTagg} MacTaggart, D., Guglielmino, S.L., \& Zuccarello F., 2016, \apjl, \textit{in press}

\bibitem[Murabito et al.(2016)]{Mur16} Murabito, M., Romano, P., Guglielmino, S.L., Zuccarello, F., \& Solanki, S. K., 2016, \apj, 825, A75

\bibitem[Reardon \& Cavallini(2008)]{Rea08} Reardon, K. P., \& Cavallini, F. 2008, A\&A, 481, 897

\bibitem[Rezaei et al.(2012)]{Rez12} Rezaei, R., Bello Gonz\'alez, N., \& Schlichenmaier, R. 2012, A\&A, 537, A19

\bibitem[Romano et al.(2013)]{Rom13} Romano, P., Frasca, D., Guglielmino, S.L., et al.\ 2013, \apj, 771, L3

\bibitem[Romano et al.(2014)]{Rom14} Romano, P., Guglielmino, S.L., Cristaldi, A. et al.\ 2014, \apj, 784, A10

\bibitem[Ruiz Cobo \& del Toro Iniesta(1992)]{SIR} Ruiz Cobo, B., \& del Toro Iniesta, J.~C.\ 1992, \apj, 398, 375 

\bibitem[Sainz Dalda \& Mart{\'{\i}}nez Pillet(2005)]{SainPil05} Sainz Dalda, A., \& Mart{\'{\i}}nez Pillet, V.\ 2005, \apj, 632, 1176 

\bibitem[Sainz Dalda \& Bellot Rubio(2008)]{SainBel08} Sainz Dalda, A., \& Bellot Rubio, L.~R.\ 2008, \aap, 481, L21
 
 \bibitem[Scherrer et al.(2012)]{Sch12} Scherrer, P.~H., Schou, J., Bush, R.~I., et al.\ 2012, \solphys, 275, 207 

\bibitem[Schlichenmaier(2002)]{Sch02} Schlichenmaier, R. 2002, AN, 323, 303

\bibitem[Schlichenmaier \& Solanki(2003)]{Sch03} Schlichenmaier, R., \& Solanki, S. K. 2003, A\&A, 411, 257

\bibitem[Schlichenmaier, Bellot Rubio \& Tritschler(2005)]{Sch05} Schlichenmaier, R., Bellot Rubio, L. R., \& Tritschler, A. 2005, AN, 326, 301

\bibitem[Schlichenmaier et al.(2010a)]{Sch10a} Schlichenmaier, R., Bello Gonz\'alez, N., Rezaei, R., \& Waldmann, T.A. 2010a, AN, 331, 563

\bibitem[Schlichenmaier et al.(2010b)]{Sch10b} Schlichenmaier, R., Rezaei, R., Bello Gonz\'alez, N., \& Waldmann, T.A. 2010b, A\&A, 512, L1

\bibitem[Schlichenmaier et al.(2012)]{Schl12} Schlichenmaier, R., Rezaei, R., \& Bello Gonz\'alez, N. 2012, in 4th Hinode Science Meeting: Unsolved Problems and Recent Insights, eds. L. Bellot Rubio, F. Reale, \& M. Carlsson, ASP Conf. Ser., 455, 61

\bibitem[Spadaro et al.(2004)]{Spad04} Spadaro, D., Billotta, S., Contarino, L., Romano, P., \& Zuccarello, F.\ 2004, \aap, 425, 309 

\bibitem[Strous \& Zwaan(1999)]{Strous99} Strous, L.~H., \& Zwaan, C.\ 1999, \apj, 527, 435

\bibitem[Tortosa-Andreu \& Moreno-Insertis(2009)]{Tort09} Tortosa-Andreu, A., \& Moreno-Insertis, F.\ 2009, \aap, 507, 949 

\bibitem[Weiss et al.(2004)]{Weiss04} Weiss, N.~O., Thomas, J.~H., Brummell, N.~H., \& Tobias, S.~M.\ 2004, \apj, 600, 1073 

\bibitem[Zhang et al.(2003)]{Zha03} Zhang, J., Solanki, S. K., Wang, J. 2003, A\&A, 399, 755 

\bibitem[Zhang et al(2007)]{Zha07} Zhang, J., Solanki, S. K., Woch, J., Wang, J. 2007, A\&A, 471, 1035

\bibitem[Zuccarello et al.(2005)]{Zucc05} Zuccarello, F., Battiato, V., Contarino, L., et al.\ 2005, \aap, 442, 661 

\end{thebibliography}
\end{document}